\documentstyle{l-aa}
\def\kms{km~s$^{-1}$}

\def\cm2{cm$^{-2}$}
\def\za{$z_{abs}$}
\def\ze{$z_{em}$}
\def\lya{Ly$\alpha$}
\def\lyb{Ly$\beta$}

\begin{document}

\newcommand{\lsim}{\ \raise -2.truept\hbox{\rlap{\hbox{$\sim$}}\raise5.truept
	\hbox{$<$}\ }}			
\newcommand{\gsim}{\ \raise -2.truept\hbox{\rlap{\hbox{$\sim$}}\raise5.truept
 	\hbox{$>$}\ }}		
\input epsf

\thesaurus{intergalactic medium -- quasars: absorption lines -- quasars:
individual: Q $0000-2619$}

\title{The shape of the ionizing UV background at $z\sim 3.7$ from the metal 
absorption systems of Q$0000-2619$
\thanks{Based on observations collected at the
European Southern Observatory, La Silla, Chile (ESO No.~2--013--49K).}}
\author{Sandra Savaglio\inst{1,2}, Stefano Cristiani\inst{3}, Sandro D'Odorico\inst{1}, 
Adriano Fontana\inst{4}, Emanuele Giallongo\inst{4}, Paolo Molaro\inst{5}}
\offprints{Sandro D'Odorico}

\institute{
	 European Southern Observatory, Garching bei M\"unchen,
        Karl-Schwarzschildstr. 2, D--85748
\and    Istituto di Astrofisica Spaziale del CNR, 00044 Frascati, Italy
\and	Dipartimento di Astronomia dell' Universit\`a, 35122
	Padova, Italy
\and    Osservatorio Astronomico di Roma, Via dell' Osservatorio 2,
        I--00040, Monteporzio, Italy
\and    Osservatorio Astronomico di Trieste, Via G.B. Tiepolo 11,
        Trieste, I--34131, Italy 
}

\date{Received date; accepted date} 

\maketitle
\markboth{The shape of the ionizing UV background from the metal 
absorption systems of Q$0000-2619$}{}

\begin{abstract} 

Spectra of the \ze~$=4.12$ quasar Q$0000-2619$ have been obtained in
the range $\lambda\lambda = 4880-8115$ \AA~with a resolution of 13
\kms~and signal--to--noise ratio of $s/n=15-60$ per resolution
element.  The list of the identified absorption lines is given
together with their fitted column densities and Doppler widths.  The
mode of the distribution of the Doppler parameters for the \lya~lines
is $\simeq 25$ \kms. The fraction of lines with $10<b<20$ \kms~is
17\%. The Doppler values derived from uncontaminated \lyb~lines are
smaller than those obtained from the corresponding Ly$\alpha$ lines,
indicating the contribution of non saturated, non resolved components
in the \lya~profiles.

The integrated UV background estimated from the proximity effect is
found to be $J \sim 7 \times 10^{-22}$ erg s$^{-1}$
cm$^{-2}$ ~Hz$^{-1}$ sr$^{-1}$. This value is consistent with
previous estimates
obtained at a lower $z$, implying no appreciable redshift evolution of
the UVB up to $z=4$.

13 metal systems are identified, five of which previously unknown.
The analysis of the associated metal systems suggests abundances
generally below the solar value with an average [C/H] $\sim -0.5$.
This value is about one order of magnitude higher than that found in
intervening systems at about the same redshift.

The analysis of the intervening metal line systems has revealed in
particular the presence of three optically thin systems with $\log
N_{HI}\sim 15$ showing associated CIV and SiIV absorptions. In order
to make the observed column densities consistent with [Si/C] ratios
lower than 10 times the solar value, it is necessary to assume a large
jump in the spectrum of the ionizing UV background beyond the HeII
edge ($J_{912}/J_{228}\gsim 1000$). This result, if confirmed in other
spectra at the same redshift is suggestive of a possible dominance of
a stellar ionizing emissivity over the declining quasar one at $z
> 3$.

\keywords{Galaxies: formation of -- general:
intergalactic medium -- quasars: Q$0000-2619$}
\end{abstract}

\section{Introduction}

\begin{table*}
\caption[t1]{Log of the observations.}
{\label{t1}}
\begin{center}
\begin{tabular}{ccccccccc}
\hline\hline&&&&&&&&\\[-5pt] 
setup & No. & date  & range & FWHM & grating & CD    & slit & exposure\\
      & of spectra && (\AA)& (\AA)&         & grism & width& (s) \\
[2pt]\hline&&&&&&&\\[-8pt] 
E1 &  2  & 15/10/90 & $4700-6452$     & 0.2 & 10 & \#5 & 1.2'' & 5400/6600 \\ 
E2 &  2  & 18/10/90 & $5830-8550$     & 0.3 & 10 & \#4 & 1.2'' & 4200/5100\\
E3 &  3  & 24-25/11/94 & $4820-8450$  & 0.3 & 10 & \#3 & 1.25''/1.25''/1.3'' & 5000/5600/5000 \\
G1 &  2  & 26-27/11/95  & $4000-10000$ & 10 & -- & \#1 & 5''/1.5'' & 900/900 \\
[2pt]\hline\\

\end{tabular}\end{center}
\end{table*} 

The \lya~forest detected in the blue side of the quasar \lya~emission
is generally ascribed to an intergalactic population of hydrogen
clouds.  \lya~clouds are present at all the observed redshifts from
the largest to the present epoch, (e.g. Carswell 1995; Bahcall et
al. 1996; Giallongo et al. 1996), covering a substantial fraction of
the total age of the universe.  The origin and evolution of the
clouds are intimately linked with the physical conditions and
evolution of the universe (Miralda-Escud\'e et al. 1996).  
While the strongest \lya~clouds showing associated metals
are thought to be associated with intervening galaxy halos (Bergeron \&
Boiss\'e 1991; Bergeron et al. 1992; Steidel et al.
 1994), the environment of the optically thin
ones is less clear.  At least at low redshift some \lya~with column
densities $\gsim 10^{14}$ cm$^{-2}$ have been associated with the
external parts of galaxy halos (Lanzetta et al. 1995).

There are however recent observational suggestions indicating a
continuity scenario between \lya~lines with $\log N_{HI}\gsim 14$ and
the stronger metal line systems.
 Clustering of the \lya~clouds has been found up to scales of 300 \kms
(Cristiani et al. 1995,1996; Chernomordik 1995; Hu et al. 1995; 
Fernandez--Soto et al. 1996) with an amplitude increasing with $N_{HI}$. In
particular, Cristiani et al. 1996 found that an extrapolation of this
trend to $\log N_{HI} \sim 17$ is consistent with the corresponding
estimate derived from the CIV metal systems by Petitjean \& Bergeron
1994.

Weak CIV lines have been found to be associated with \lya~lines with $\log
N_{HI} = 14.5 \sim 15$ (Cowie et al. 1995; Tytler et al. 1995) with
abundances similar to that of the metal line systems.

Disentangling between different scenarios for the cloud structure and
their cosmological evolution requires a large database of high resolution
spectra. The sample available in literature is still limited,
but new impulse in the field has come from the observations
with the HIRES spectrograph at the Keck telescope 
(Cowie et al. 1995; Hu et al. 1995)

Within an ESO Key Project on intergalactic matter at high redshift, we have obtained
at the 3.5m NTT high resolution spectra of several QSOs at redshift larger than 3
(Giallongo et al. 1996). We present and discuss here in detail data on the \lya~forest 
and the metal systems of the QSO $0000-26$ ($m_R = 17.5$, $z_{em} = 4.12$).
The spectra cover the range 4880 to 8115 \AA~at a resolution of 13 \kms.
A preliminary investigation of the spectrum of Q$0000-26$ in the range
4700 to 6600 {\AA} at a resolution of 30 \kms~ has been reported
by Webb et al. 1988.  A first discussion of the
metal systems in Q$0000-26$, based mainly on the data in the spectral
region around the Ly$\alpha$ emission and to the red of it, has been
given by Savaglio et al. 1994, while a detailed study of the metallicity
of the damped \lya~system at $z=3.39$ has been reported by Molaro et
al. 1995. 

In this paper we present in section 2 the observations, the data reduction procedure
and the list of the \lya~lines and identified metal lines with the fitting parameters
column density, Doppler width and redshift. The result of the statistical analysis
of the \lya~forest is presented in section 3. The metal line systems are discussed in
section 4.

\section{Observations and  data extraction}

\begin{figure}
\caption[1]{\label{f0} Signal--to--noise ratio per resolution element
as function of wavelength in the spectrum of Q$0000-26$.}
\epsfxsize=9cm
\epsfysize=9cm
\epsffile{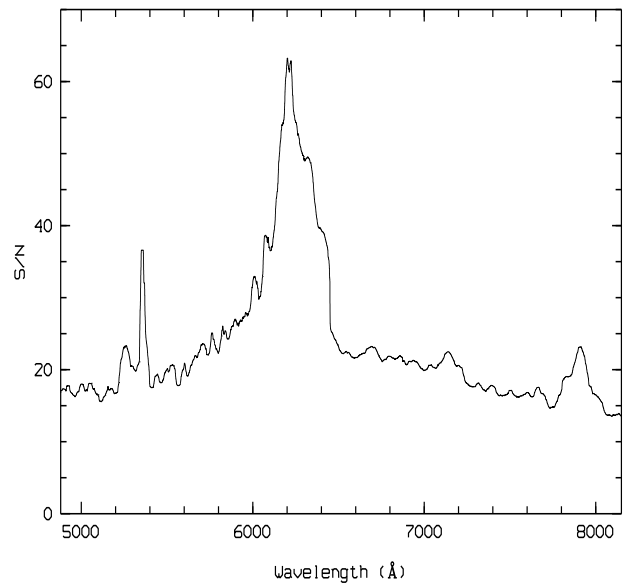}
\end{figure}

\begin{figure*}
\caption[1]{\label{f1a} Absorption lines of Q$0000-26$. The 
normalized spectrum is shown as full line, the fit as thin line and the
noise per pixel as dashed line. Long ticks correspond to \lya~and
\lyb~lines, short ticks correspond to  metal lines. For $\lambda < 5260$ \AA~
the \lyb~lines are indicated with the same number as the corresponding
\lya~lines.}
\epsfxsize=19cm
\epsfysize=23.5cm
\epsffile{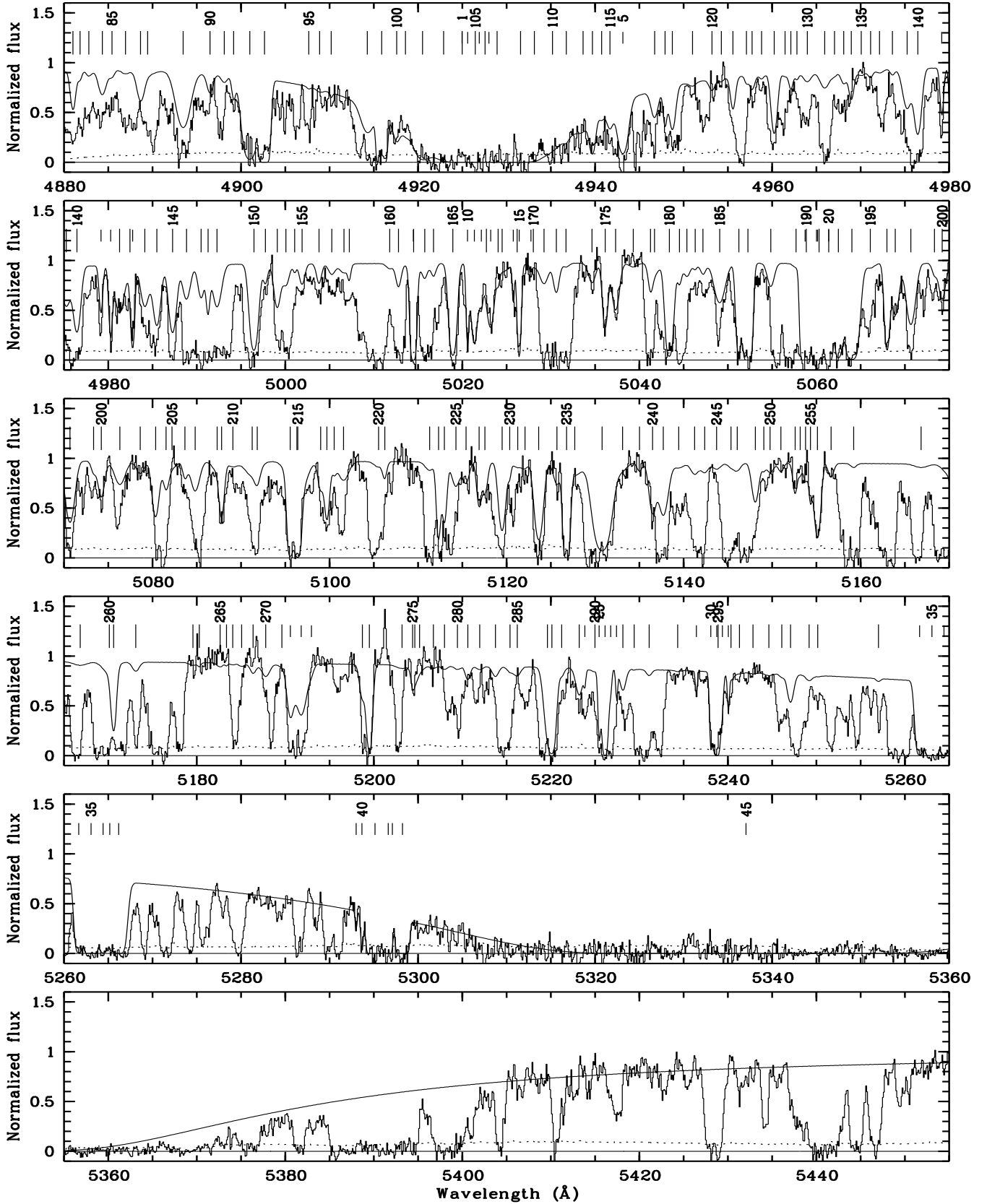}
\end{figure*}

\setcounter{figure}{1}
\begin{figure*}
\caption[1]{\label{f1b} {\it -- Continued}}
\epsfxsize=19cm
\epsfysize=23.5cm
\epsffile{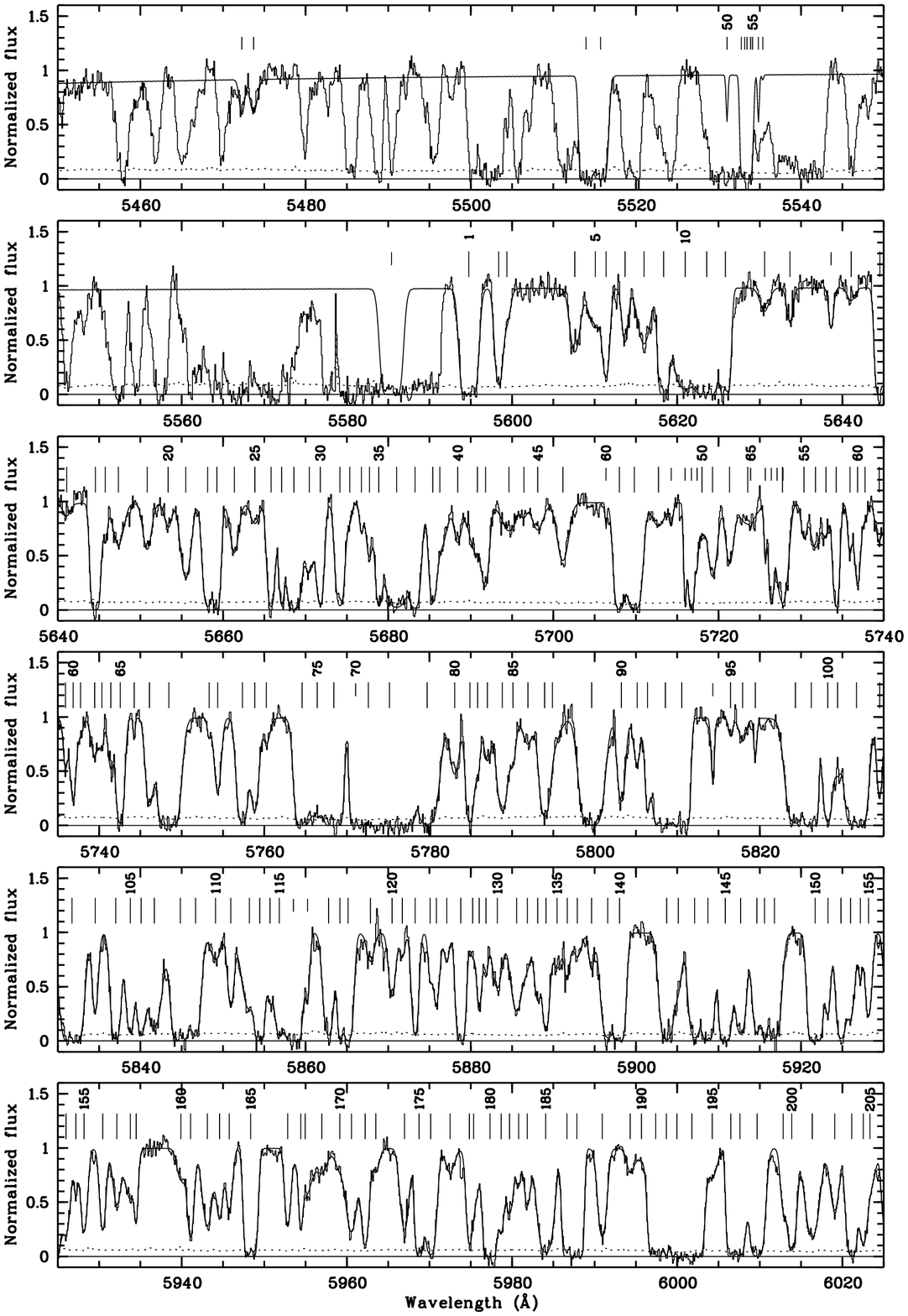}
\end{figure*}

\setcounter{figure}{1}
\begin{figure*}
\caption[1]{\label{f1c} {\it -- Continued}}
\epsfxsize=19cm
\epsfysize=23.5cm
\epsffile{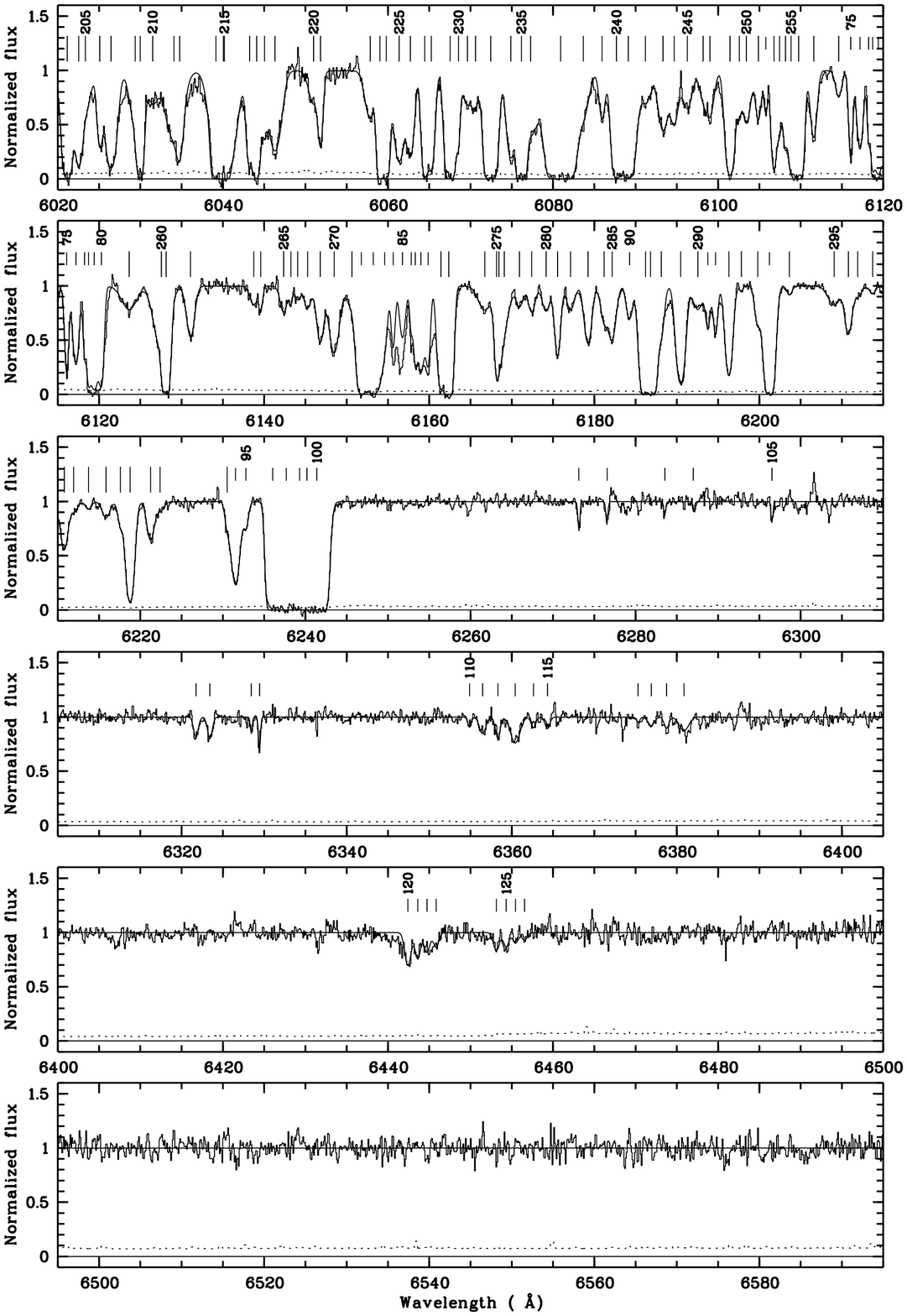}
\end{figure*}

\setcounter{figure}{1}
\begin{figure*}
\caption[1]{\label{f1d} {\it -- Continued}}
\epsfxsize=19cm
\epsfysize=23.5cm
\epsffile{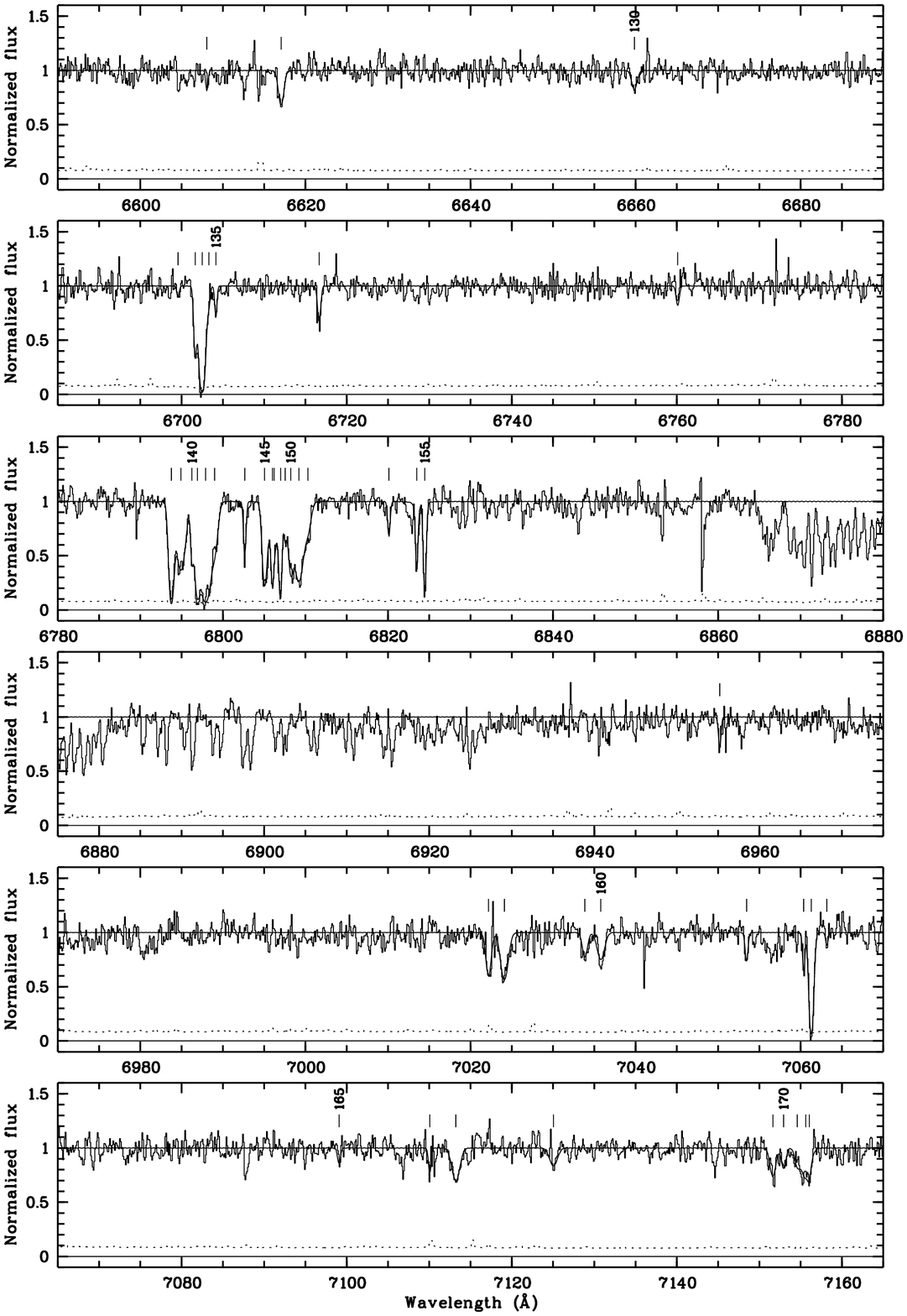}
\end{figure*}

\setcounter{figure}{1}
\begin{figure*}
\caption[1]{\label{f1e} {\it -- Continued}}
\epsfxsize=19cm
\epsfysize=23.5cm
\epsffile{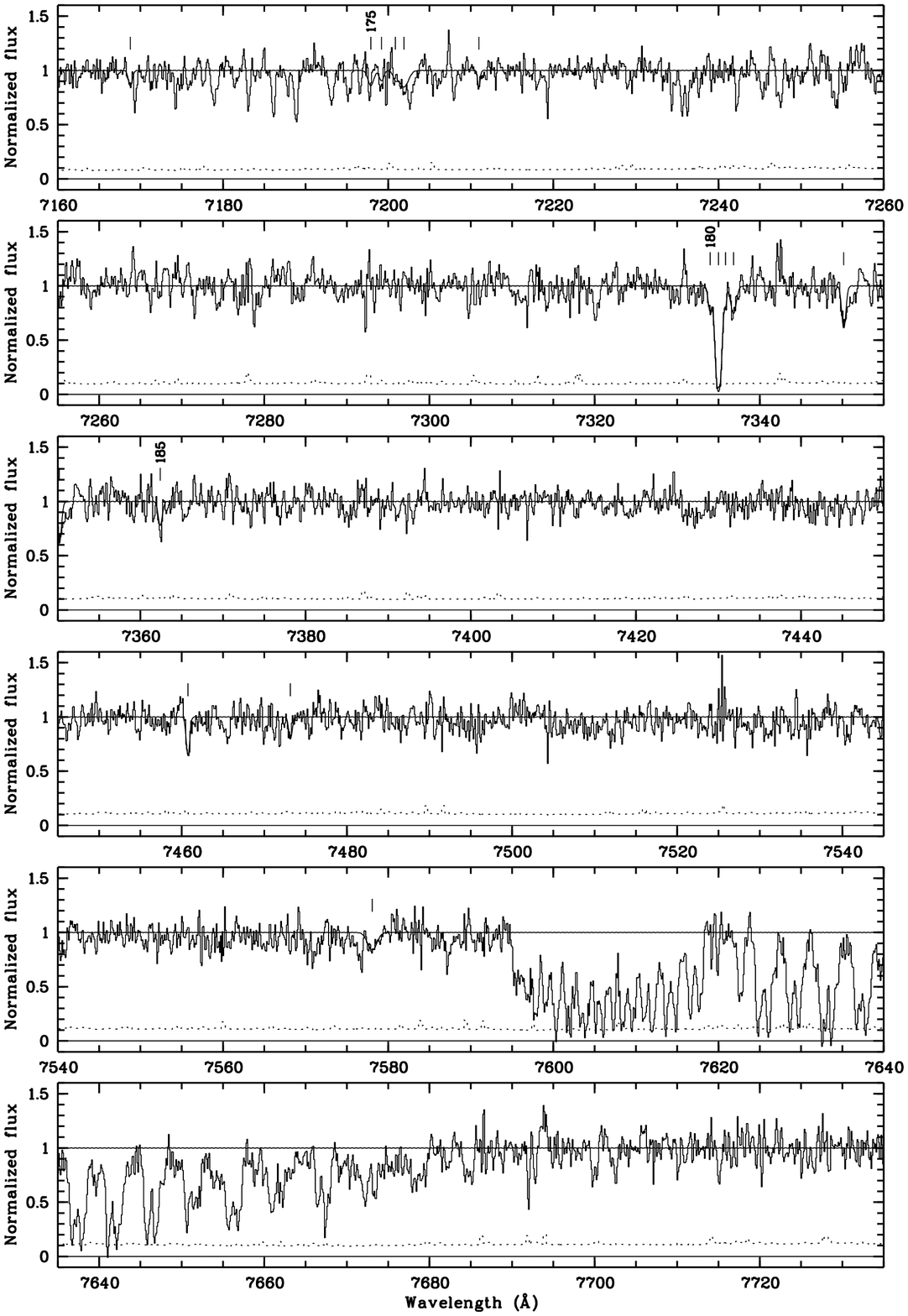}
\end{figure*}

\setcounter{figure}{1}
\begin{figure*}[t]
\caption[1]{\label{f1f} {\it -- Continued}}
\epsfxsize=19cm
\epsfysize=16.7cm
\epsffile{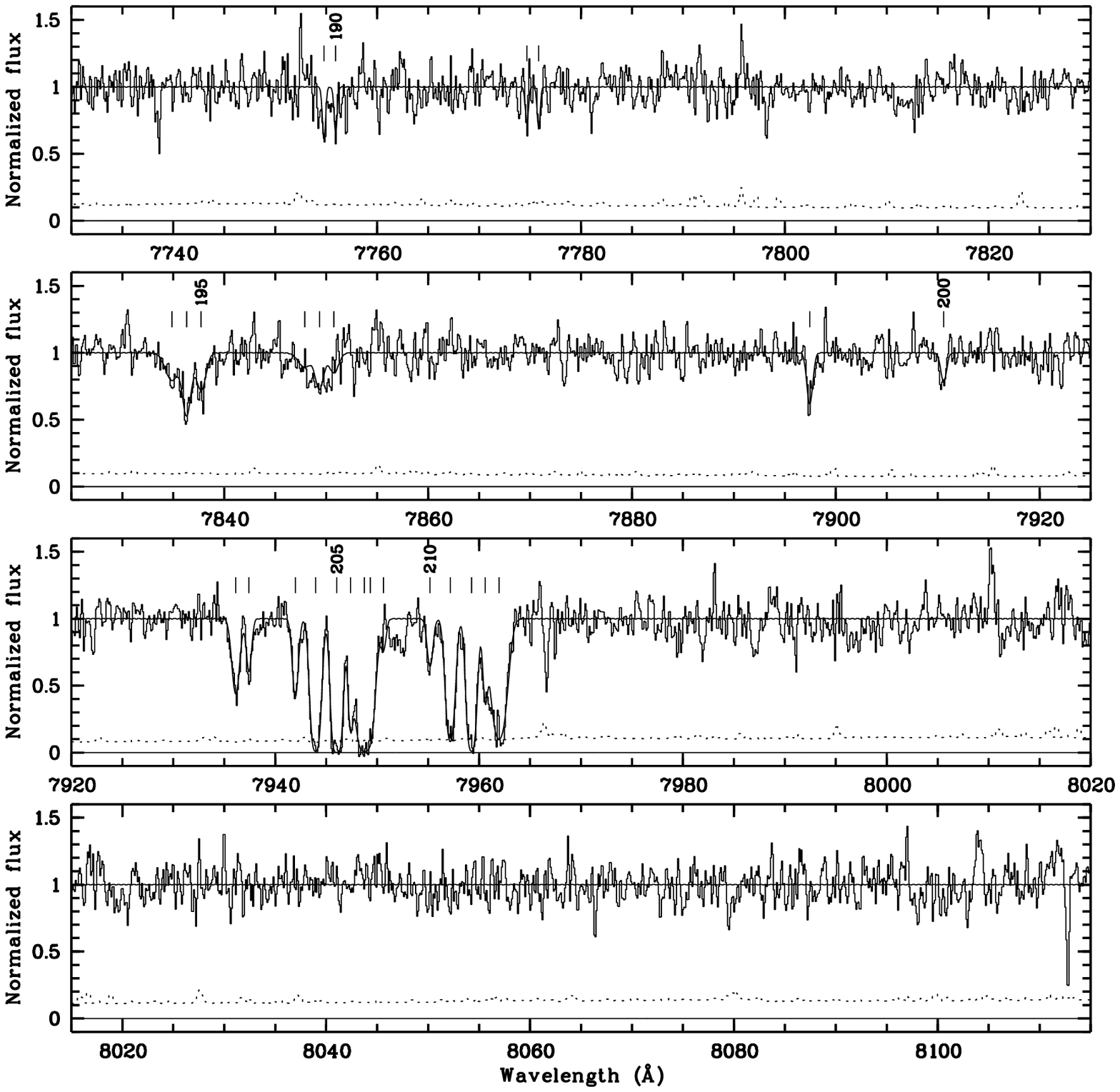}
\end{figure*}

The echelle observations of Q$0000-2619$ presented here were obtained with the
EMMI instrument (D'Odorico 1990) at the ESO NTT telescope in October
1990 and November 1994. 
The log of the observations is given in Table 1.  
The second column gives the number of individual spectra obtained
with each setting.  The seeing during the observations was typically
between 0.8 and 1.2 arcsec.

The absolute flux calibration was carried out by observing of the 
standard stars Feige 110 (Stone 1977), LTT7987 (Stone
and Baldwin 1983 and Stone and Baldwin 1984) and HD49798 (Walsh
1992).

The echelle data were reduced with the ECHELLE software package available in
the MIDAS software. The wavelength calibration spectra of the
Thorium-Argon lamp were extracted in the same way and used to
establish the wavelength scale.  Wavelengths have been corrected to
vacuum heliocentric values.  The weighted mean of the spectra has been
obtained at the resolution of about $R=24000$. The variance spectrum was
obtained by propagation of the photon statistics of the object and sky
spectra, and from the detector read--out--noise.
The final signal--to--noise ratio per resolution element is shown in 
Fig.~\ref{f0}.

The normalized spectrum is plotted in Fig.~\ref{f1a}
in the wavelength interval $\lambda\lambda =4880-8115$ {\AA}. 
The dedicated  software FITLYMAN (Fontana \&
Ballester 1995) in the MIDAS package was used to derive the redshift
$z$, the Doppler parameter $b=\sqrt{2} \sigma$ and the column density
$N$ of the absorption lines.  The line fitting has been performed by a
$\chi^2$ minimization of Voigt profiles, after deconvolution with the
instrumental profile.  

Despite the high resolution, most of the features appears to be
strongly blended, contrary
to what is found in lower $z$ QSOs, where the
lines are typically isolated. As  in previous similar analyses
(e.g.~Giallongo et al. 1993), complex structures
have been fitted with the minimum
number of components required to give a probability of random
deviation $P>0.05$.

We performed the fitting of all the line complexes in the region 
with $s/n\gsim 20$, i.e.  from $z= 3.602$,
to the quasar \lya~emission line ($\lambda\lambda = 5590-6240$ \AA).  

The parameter list for about 300 \lya~lines is reported in Table \ref{t2},
while Table \ref{t2b} lists the metal lines. 
Only the fitted lines appear in the tables.

The position of fitted lines is marked on the
top of Fig.~\ref{f1a} with the numbering as given in Table \ref{t2}
and \ref{t2b}. Long ticks show \lya~lines (Table \ref{t2}) 
and with the same numbering
the associated \lyb~lines in the wavelength range $4880-5260$ \AA.
Short ticks are  metal
absorption lines (Table \ref{t2b}).

The \lya~forest is contaminated by the metal lines of two damped
\lya~systems and other 11 metal systems, 4 of which with \za~$\approx$
\ze.  In Table \ref{t2b} the
\lya~lines associated with metal systems
are indicated as ``M\lya" and taken off from
the sample used for \lya~analysis and statistics.

A low resolution ($R=280$) spectrum of Q$0000-26$, covering the
range $\lambda \lambda = 4000-10000$ \AA, has been obtained in the long
slit mode of EMMI at the NTT in November 1995 (see Table 1). 
The absolute flux
calibration was carried out observing the standard star Feige 110. 

\section{The \lya~forest}\label{s4}

Clues on the physical nature of \lya~clouds may be obtained
from the statistical distributions of their parameters (redshift, 
Doppler width and column density) obtained through line profile fitting.

The present data,  covering a wide redshift range with
good signal--to--noise, are especially suited to address the issue
of the Doppler parameter distribution and to study the UV background
at $z\sim 3.5-3.8$.
The redshift evolution and column density distribution of the
\lya~clouds have been already discussed using a larger data base 
in Giallongo et al. (1996).

\subsection{The Doppler parameter distribution}\label{s4.1}

While it is generally agreed that  ``typical'' \lya~clouds
have Doppler parameters of the order of $b\sim 25$ \kms,
the actual fraction of narrow ($b<20$ \kms) and broad ($b>50$ \kms)  lines 
is more difficult to determine, because of  the possible
systematic effects involved.
 
Large $b$ values may be a result
of the intrinsic difficulty in finding out sub-components in
blends, while noise effects and the contamination 
of unrecognized metal lines may increase the fraction of narrow lines
(Rauch et al. 1993).

\begin{figure} 
\caption[1]{\label{f4a} The Doppler parameter $b$ as function of wavelength.}
\epsfxsize=9.5cm 
\epsfysize=9.5cm 
\epsffile{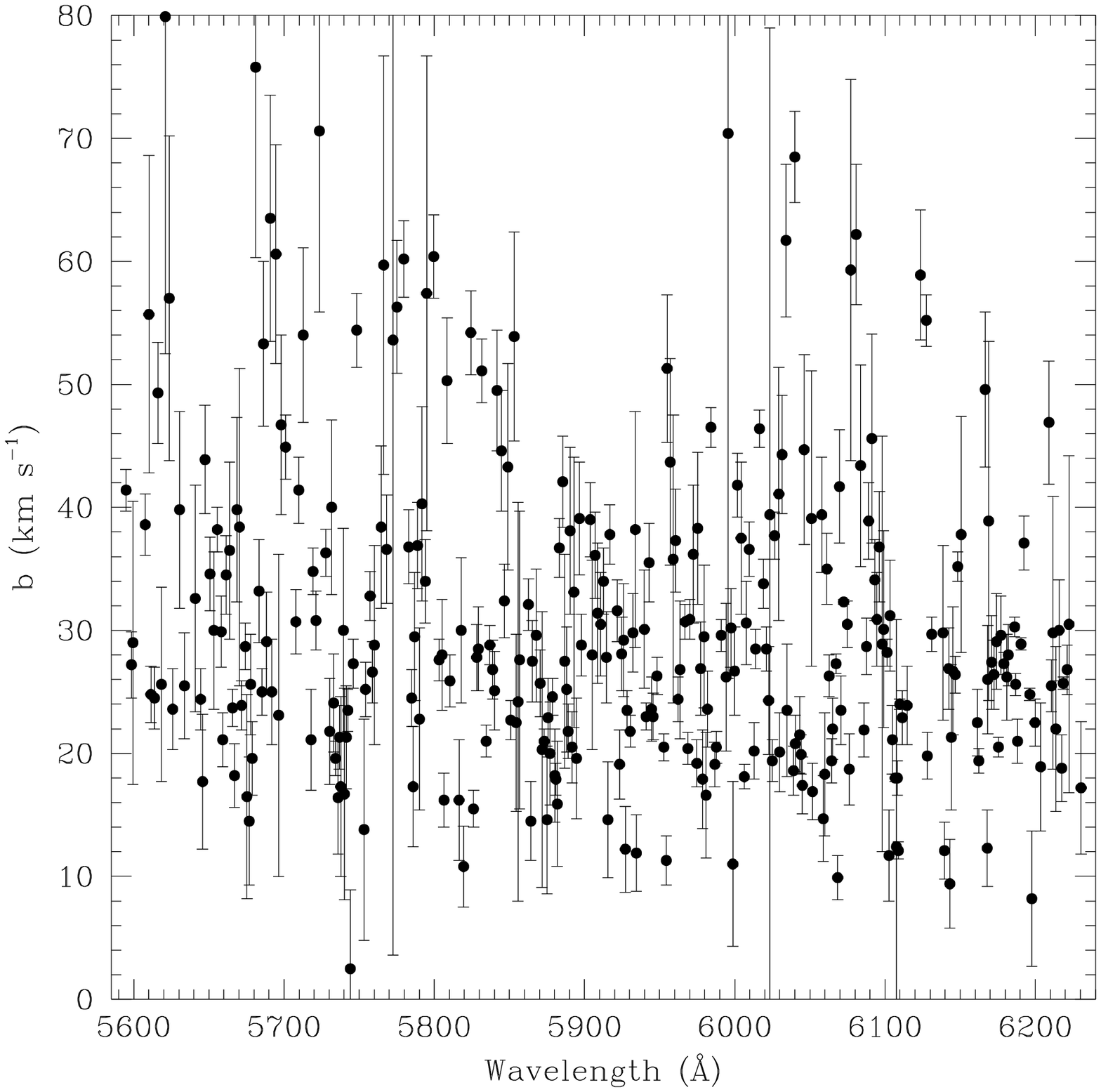} 
\end{figure}

The strong biases in the detection and measure of the narrow
lines can be minimized with high quality data on an extended 
redshift range, as in the present spectrum.
As shown in Fig.~\ref{f4a}, low $b$ values are not correlated with
the wavelength and consequently with the $s/n$ (see also Fig.~\ref{f0}).

\begin{figure} 
\caption[1]{\label{f2b} The Doppler parameter distribution, with a median 
value of about 27.8 \kms. The lines are those not affected by the
proximity effect.} 
\epsfxsize=8cm 
\epsfysize=8cm 
\epsffile{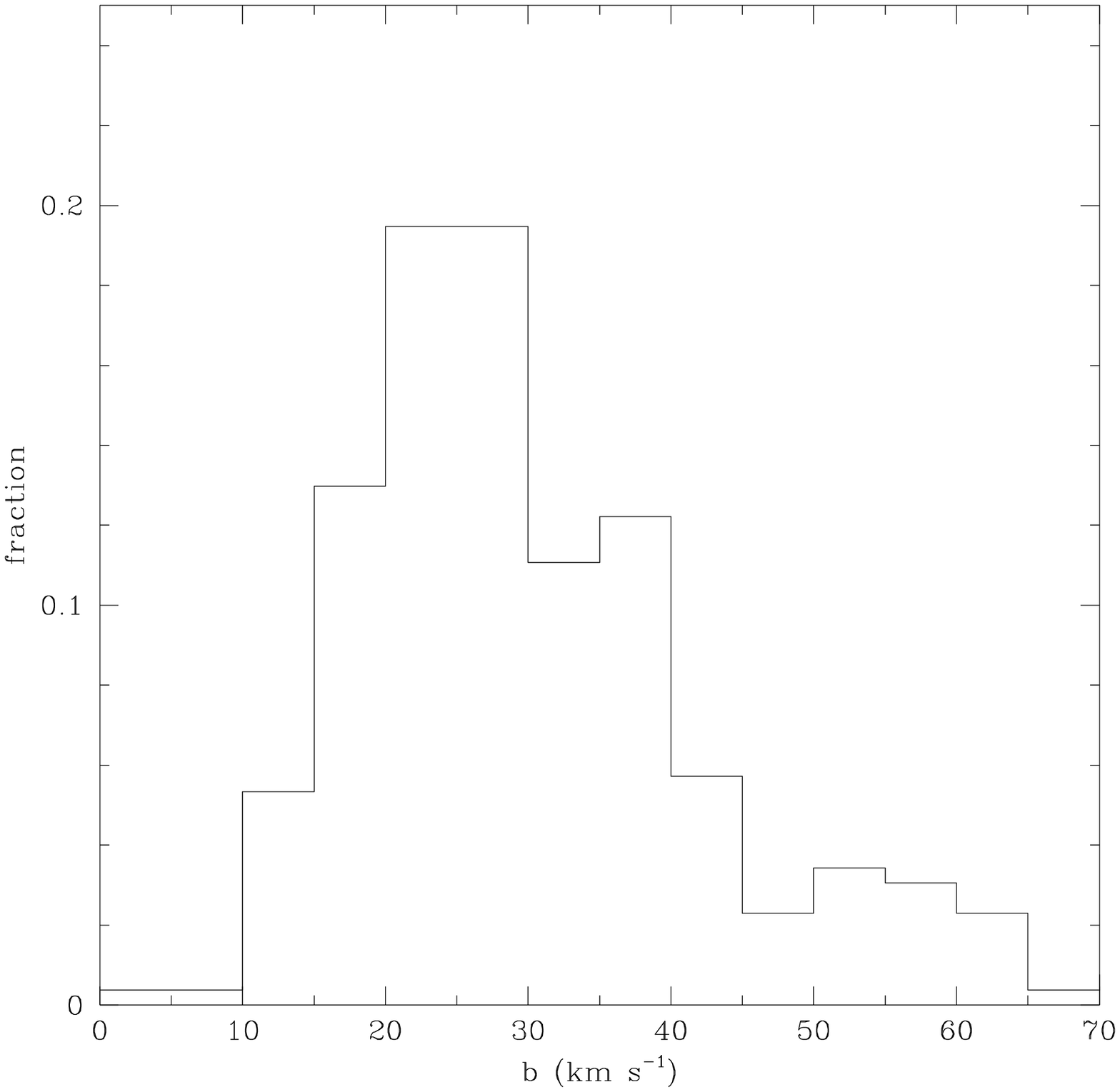} 
\end{figure}

The Doppler parameter distribution has been obtained 
by selecting all the lines out of 8
Mpc from the QSO, not affected by the proximity effect, and
is shown in Fig.~\ref{f2b}.

As usual, the distribution appears skewed towards large $b$ values.
The mode of the distribution 
is 25 \kms~and 17\% of the lines have $10 < b < 20$.  

To estimate the intrinsic dispersion of the distribution we calculate
iteratively the mean $b$ value excluding lines with 2$\sigma$
beyond the mean. In this way we avoid large, possible spurious $b$
deriving a mean value $\langle b
\rangle = 26$ \kms~and $\sigma _b=8$ \kms.  The
distribution of measurement errors has a median value $\simeq 3$ \kms~so
does not affect appreciably the observed $b$ dispersion.
 After subtraction in quadrature we obtain $\sigma _b=7$ \kms.

Of course any
intrinsic $b$ distribution with an artificial cutoff at the low end
produces a low$-b$ tail due to measurement errors (Hu et al. 1995).  A
very large statistics with low measurement errors is needed to
deconvolve the intrinsic distribution from the observed one. Without a
very large \lya~sample, the problem of the intrinsic fraction of
narrow lines remains an open question although photoionization models
(Giallongo \& Petitjean 1994; Ferrara \& Giallongo 1996) and
recent cosmological models for the \lya~clouds (Hernquist et al. 1996;
Miralda--Escud\'e et al. 1996) are able to produce $b$ values as
low as $b\sim 15$ \kms.

\subsection{The \lyb~forest and the $b$ values}\label{s4.3}

\begin{table*}\caption[t2]{Absorption line parameters of the
\lya~forest. Lines marked with an asterisk have been fitted using 
simultaneously the observed \lya~and \lyb~profile.}
{\label{t2}}
\epsfysize=26cm
\epsffile{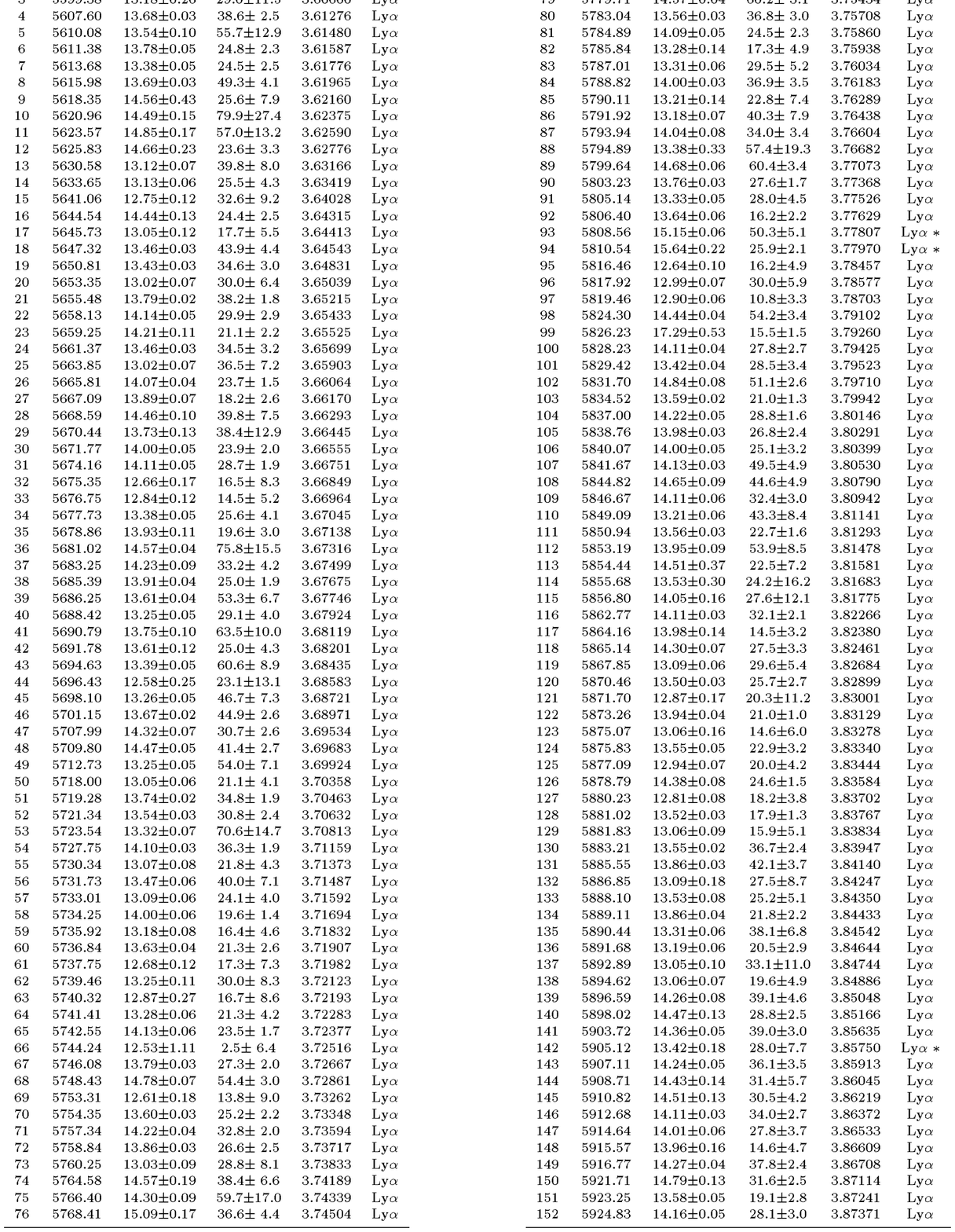}
\end{table*}

\setcounter{table}{1}
\begin{table*}\caption[t2a]{\it -- Continued}
{\label{t2a}}
\epsfysize=26cm
\epsffile{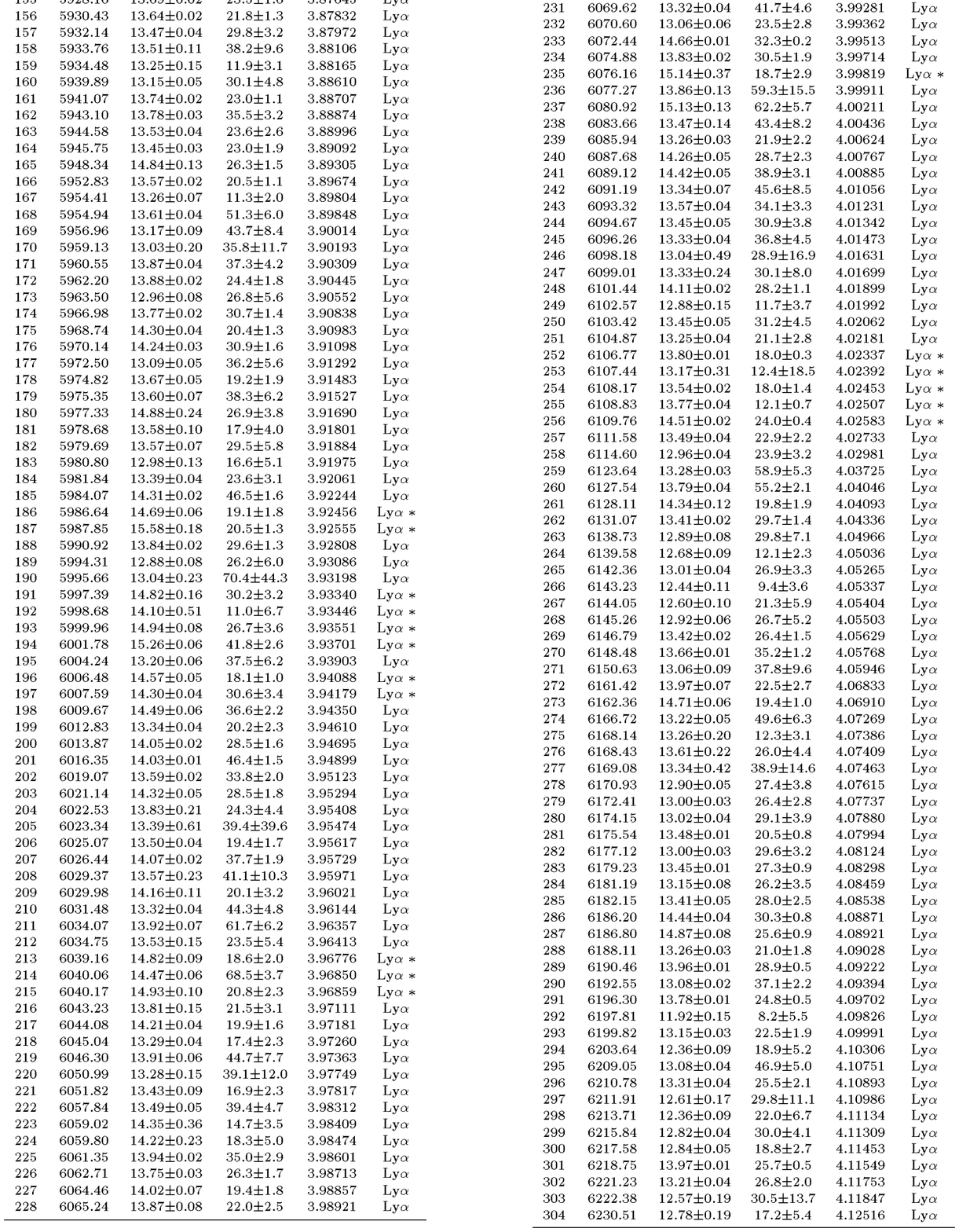}
\end{table*}

\begin{table*}\caption[t2b]{Absorption line parameters of the metal systems.}
{\label{t2b}}
\epsfysize=26cm
\epsffile{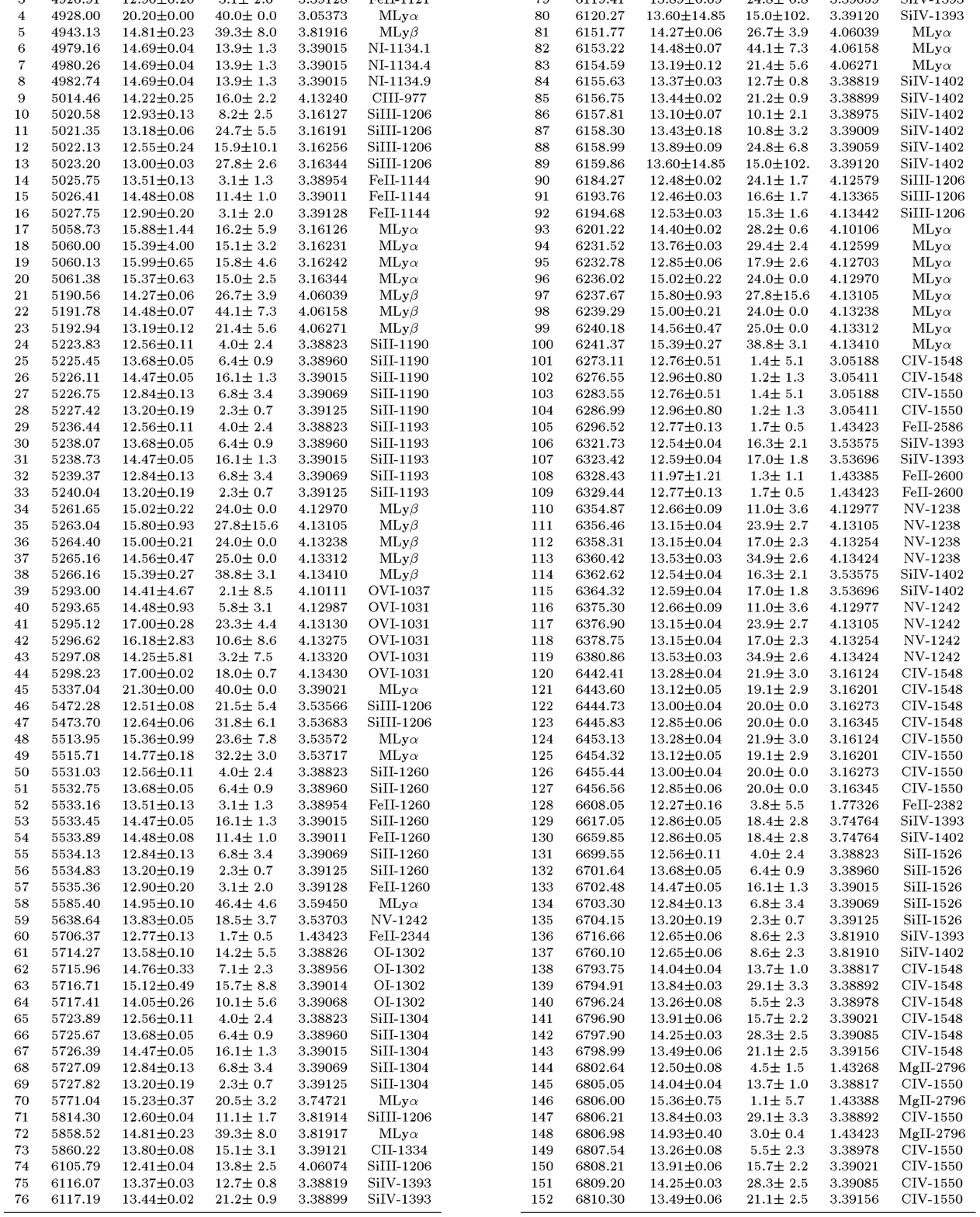}
\end{table*}

\setcounter{table}{2}
\begin{table}\caption[t2c]{\it -- Continued}
{\label{t2c}}
\epsfysize=26cm
\epsffile{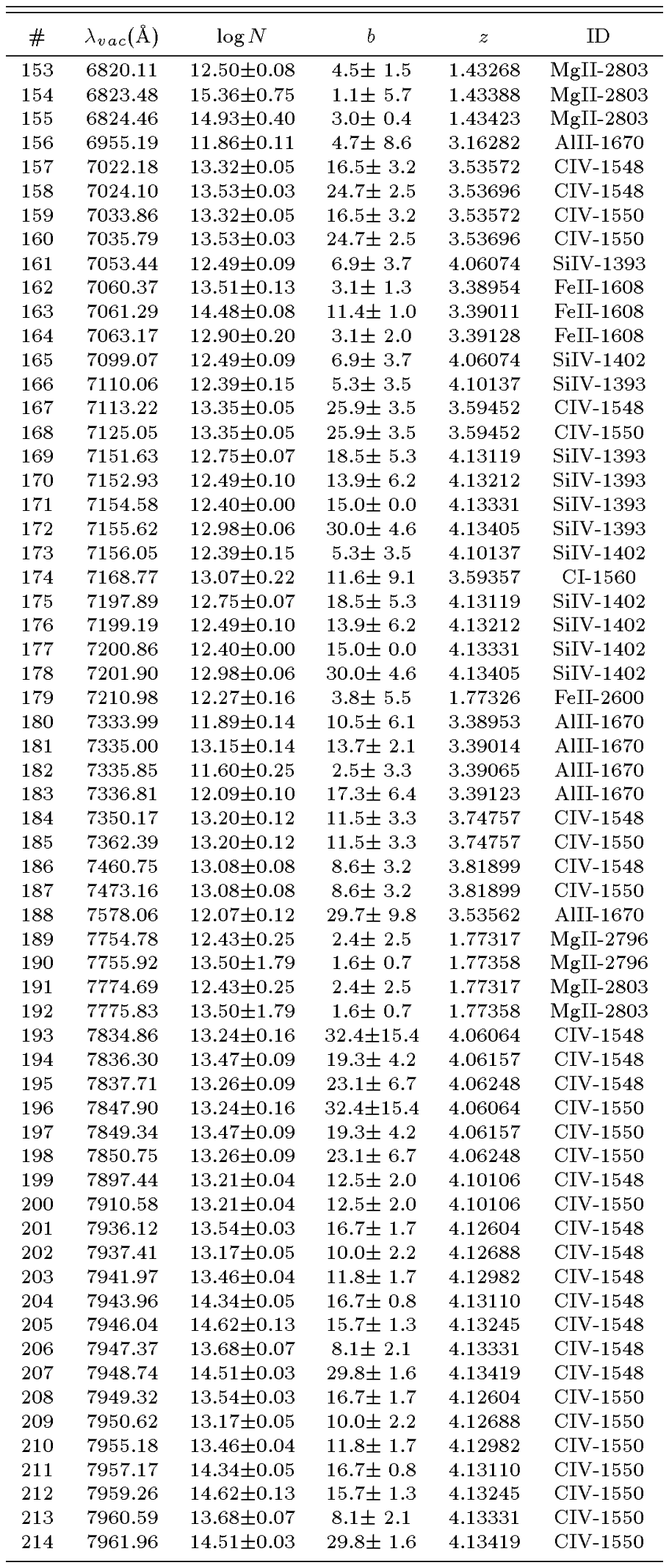}
\end{table}

\begin{figure} 
\caption[1]{\label{f3} Plot of the Doppler parameter $b$ versus $\log
N_{HI}$ for a subsample of absorption lines. Solid dots are parameters
obtained from a fit of the \lya~only, the squares show the result of the
fit when \lyb~is also considered. One can note a splitting of the
higher column density lines and a lowering of the $b$ values. }
\epsfxsize=9cm 
\epsfysize=9cm 
\epsffile{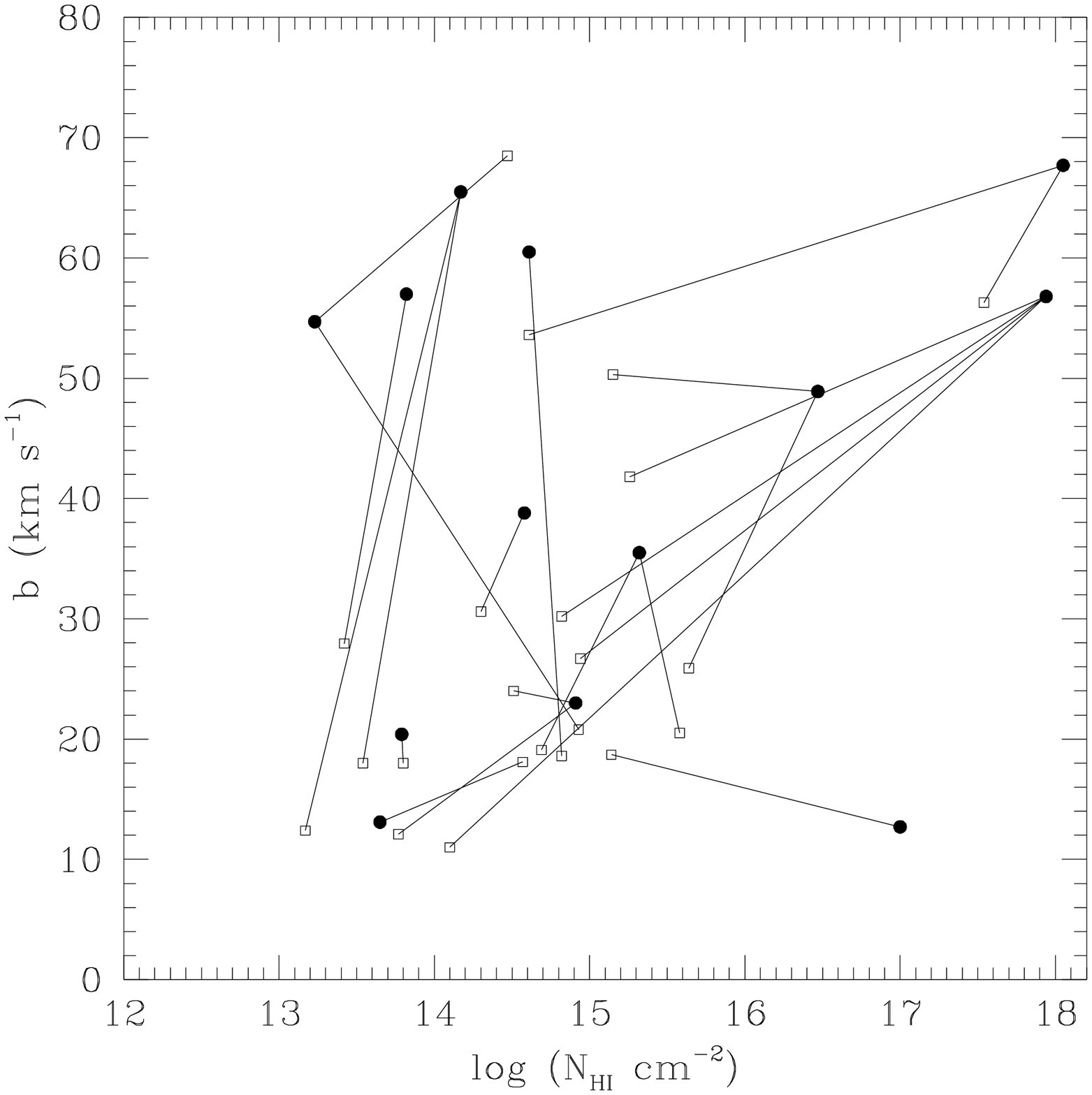} 
\end{figure}

The line parameters derived from \lya~fitting have been compared with
the corresponding \lyb~lines.  In general the \lyb~forest
is mixed with the \lya~forest at lower $z$, but in several cases we
found isolated absorptions in correspondence of the position of the
expected \lyb.  
For these systems we tried a simultaneous fit for the \lya~and \lyb~
components of the same cloud. Since the \lyb~line in these cases is
less saturated than the corresponding \lya, a more accurate estimate
of $N_{HI}$ and $b$ is obtained.  In Fig.~\ref{f3} we show
a sub-sample  for which the \lya+\lyb~fit  gives
a significantly different
result from the fit with only 
the \lya~component. 
In general the high column density \lya~lines tend to split
in more components: the initial  13 lines become  22 after decomposition.
These 22 lines are marked with an asterisk in Table \ref{t2}.
Besides the $b$ mean value goes from 43 \kms~to 28 \kms
~in better agreement  with the $b$ distributions derived in 
Sect 3.1 and by Hu et al. (1995).

While a firm statistical conclusion cannot be drawn  with
these few cases, they suggest that at least some of the
lines with large $b$  are due to
blends of several components. 
In this way the tendency  of lines with larger $b$  to show
larger column densities is strongly reduced.

\subsection{The UV background at $z\lsim 4$}\label{s5.3}

\begin{figure}
\caption[1]{\label{f5} $\partial n / \partial X_\gamma$ as function of
$\log \omega$ for lines with $ \log N_{HI} \geq 13.8$ and for the UV background
radiation $J \simeq 7 \times 10^{-22}$ erg s$^{-1}$ \cm2~Hz$^{-1}$
sr$^{-1}$.}
\epsfxsize=8cm 
\epsfysize=8cm
\epsffile{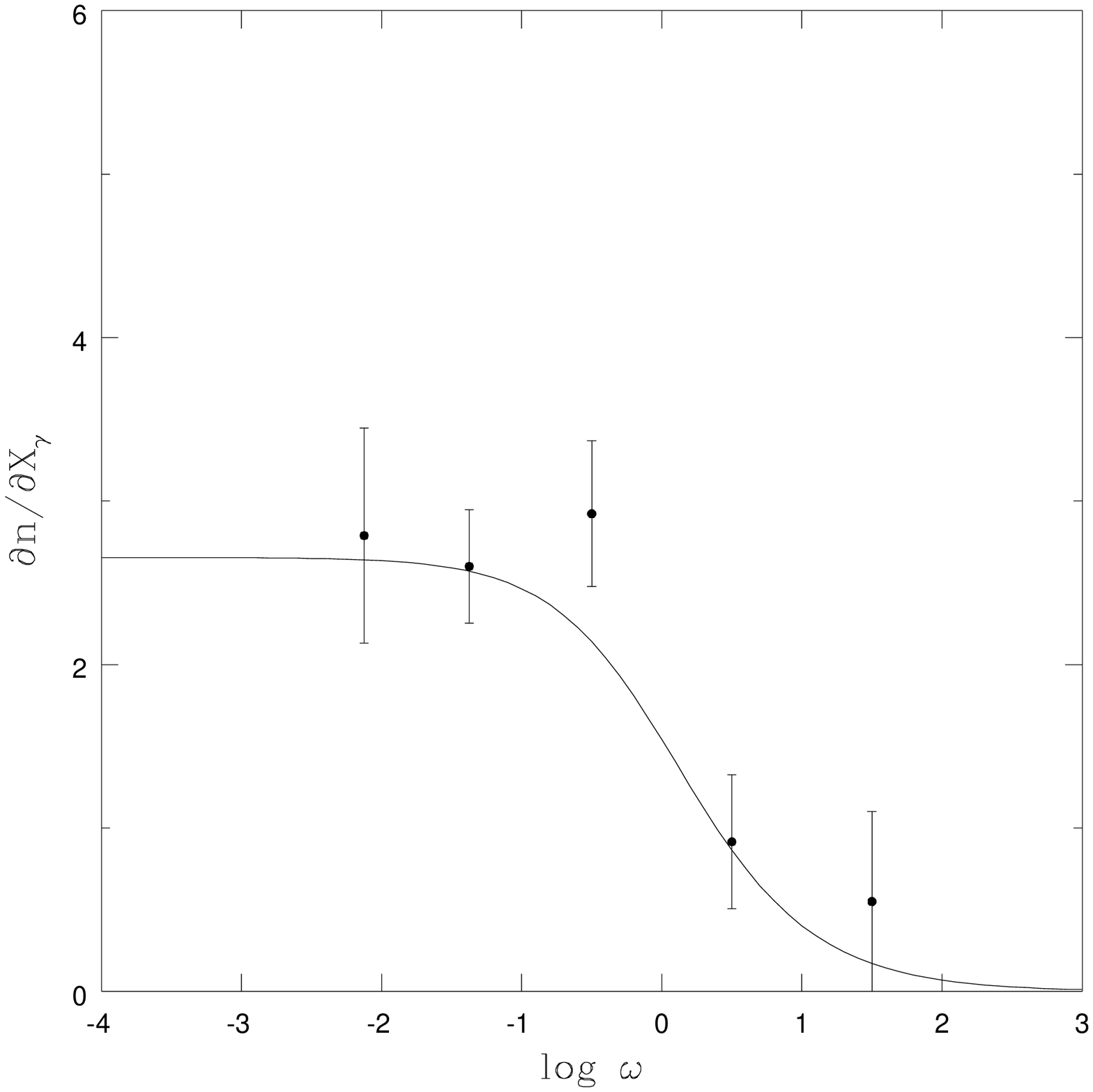}
\end{figure}

The reduction of the number density of the absorption lines along the
wing of the quasar \lya~emission is interpreted as due to the
enhancement of the ionization of the gas cloud by the UV emission
of the nearby quasar which is superimposed to the general UV background
(Bajtlik et al. 1988).  This proximity effect allows a
statistical estimate of the UVB as a function of redshift.

Following the Bajtlik et al. model, the line distribution per unit
column density can be represented in the proximity of the QSO by
(the subscript HI is omitted for simplicity):

\begin{equation}\label{eq5}
\frac{\partial ^2 n}{\partial z \partial N} = A(1+z)^{\gamma}N^{-\beta}
(1+\omega)^{(1-\beta)} ~.
\end{equation}
where $\omega$ is the ratio of the quasar Lyman limit flux to the 
background flux received by any cloud at its redshift.

Assuming a power law spectrum $f_\nu \propto \nu^{-\alpha}$, with 
$\alpha=0.72$ (Schneider et al.
1989) and the continuum flux estimated at the minimum between SiIV and CIV
emissions from the spectrum of Schneider et al. 1989, the flux at 912 \AA~ is
$f_{912} = 2.5 \times 10^{-27}$ erg s$^{-1}$ \cm2~Hz$^{-1}$.
Uncertainties in the calculation of $\omega$ depend mainly on the
estimate of the systemic emission redshift of the quasar. As shown by  Espey
et al. (1993), the best estimate of  the actual redshift is given by the low
ionization lines, as for example the MgII doublet.  
We adopted the value $z_{em} = 4.126$ as we derived from the fit to the OI(1302)
emission line.

We have considered the sample of \lya~lines not associated with metal
systems with $z\geq 3.60$ and $\log N_{HI} \geq 13.8$. The high threshold
adopted for the column density of the sample avoids the bias in the
redshift distribution of the weaker lines due to the blanketing effect
produced by the stronger ones.

In Fig.~\ref{f5} the predicted distribution of the line density in the
coevolving redshift interval $dX_{\gamma}=(1+z)^{\gamma}$ is shown
together with the data points binned as in Bajtlik et al. (1988). 
The best fit gives $J = J_{-22} \times 10^{-22}$ erg s$^{-1}$ 
\cm2~Hz$^{-1}$ sr$^{-1}$ with $J_{-22} \simeq 7$
assuming an intrinsic line distribution with $\gamma
=2.89$, $\beta = 1.79$ and $A=10^{11.22}$, as derived from the statistical
analysis of the data sample used by  Giallongo et al. (1996)
for lines with $\log N_{HI} \geq 13.8$. 
Only values $J_{-22}<4$ and $J_{-22}>18$ are excluded at more than
2$\sigma$ level. This result is consistent with the value $J_{-22}=6$
derived from a large data
sample (including this spectrum) by Giallongo et al. (1996) 
without correction for line blanketing. Including line blanketing corrections
in the large sample reduces the UVB from $J_{-22}=6$ to $J_{-22}=5$.
At higher redshift, $z\sim 4.5$, Williger et al. (1994), 
have measured the proximity effect in the forest of BR$1033-03$ 
obtaining a value $J_{-22}\sim 3$. This might imply a possible
evolution of the UVB at $z>4$, which is to be confirmed by a larger
data set.

The value derived from the proximity effect in 
Q$0000-26$ is not far from that
predicted for the quasar population at the same redshift $J_{-22} \sim 1-2$
(Haardt \& Madau 1996),
although there is room for a contribution by other kind of ionizing
sources like primeval galaxies. 

Disentangling between these two possibilities requires the knowledge
of the shape of the UV background around the HeII edge at 4 Rydberg
(228 {\AA}). This can be done either through the direct measure of the
quasar flux at 4 $Ryd$ in the few cases where the quasar spectrum can be
observed in this region (Jakobsen et al. 1994; Davidsen et al. 1995)
or in an indirect way through the measure of the relative abundances
of ions like CIV and SiIV whose ionization potentials are near the
HeII edge (Miralda-Escud\'e \& Ostriker 1990).  In the next section we
derive constraints on the shape of the UVB and on the nature of
the ionizing sources from the study of three optically thin \lya~
absorption systems at $z\gsim 3.5$.

\section{The metal systems}

The metal systems of $0000-26$ have been already studied
in Savaglio et al. (1994). 
In this work, the new data allow to confirm the old metal
systems (except one) and to identify five new ones, with relatively low HI
column density. 
Table \ref{t3} lists the two low redshift metal
systems containing the MgII absorption doublet.
Table \ref{t4a} shows the CIV high-redshift systems
with $z_{em}-z_{abs} > 5000$ \kms, considered to be intervening.
Table \ref{t4b} lists the CIV systems with $z_{em}-z_{abs} < 5000$
\kms, considered to be  associated.
All the high-redshift systems show CIV doublet together with the \lya~line.

For all the systems we looked for metal lines of cosmological
relevance falling in the observed range. For most of these, we give
upper limits to the column density assuming a $b$ value as reported in the
Tables.
Statistical errors of the line parameters are given 
as a result of the line fitting procedure adopted. 

\subsection{Ionization in the intervening systems}

\begin{table*}\caption[t1]{Low redshift metal systems of Q$0000-26$.
Upper limits to column densities are obtained for $b$ values between 
brackets.}
{\label{t3}}
\begin{center}
\begin{tabular}
{|l|cc|cc|cc|}
\hline\hline&&&&&&\\[-5pt]
\multicolumn{1}{|c|}{$z_{abs}$}
& \multicolumn{2}{c|}{FeII} 
& \multicolumn{2}{c|}{MgII} 
& \multicolumn{2}{c|}{MgI} \\
[2pt] & $\log N$ & $b$  & $\log N$ & $b$  & $\log N$ & $b$ \\
\hline\hline&&&&&& \\[-5pt]
1.4326 &  $13.47\pm1.47$ & $0.9\pm0.4$  &  $12.50\pm0.08$  & $4.5\pm1.5$ &  $<12
.0$  & (5) \\
1.4342 &  $12.77\pm0.13$ & $1.7\pm0.5$  &  $14.93\pm0.40$  & $3.0\pm0.4$ &  $<11
.7$  & (5) \\
1.4338 &  $11.97\pm1.21$ & $1.3\pm1.1$  & $15.36\pm0.75$    & $1.1\pm5.7$ &  $<1
1.4$  & (5) \\
[4pt]\hline&&&&&&\\[-8pt]
1.7732  & $<12.7$    & (10)  & $12.43\pm0.25$ &  $2.4\pm2.5$ & $<11.6$&(10)\\
1.7736  & $<12.5$    & (10)  & $13.50\pm1.79$ &  $1.6\pm0.7$ & $<11.6$&(10)\\
[4pt]\hline\end{tabular}\end{center}
\end{table*} 

\begin{table*}\caption[t2]{Intervening metal systems.
Upper limits to column densities are obtained for $b$ values between 
brackets. Detections reported with an asterisk are doubtful
because in the \lya~forest or very weak. The reported redshifts are
referred to the CIV doublet.}
{\label{t4a}}
\epsfysize=26cm
\epsffile{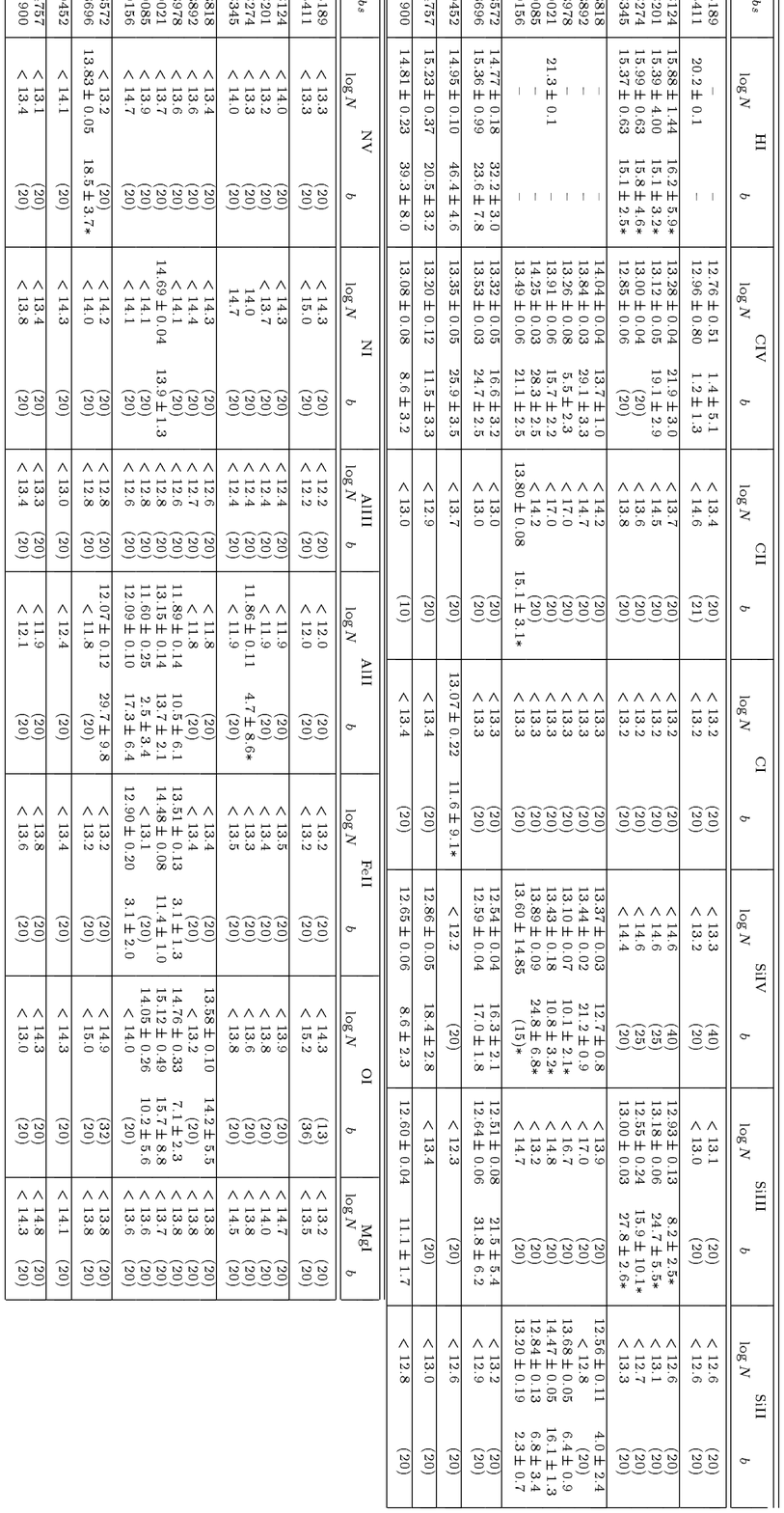}
\end{table*}

\begin{table*}\caption[t2]{Associated metal systems.
Upper limits to column densities are obtained for $b$ values between 
brackets. Detections reported with an asterisk are doubtful
because in the \lya~forest or very weak. The reported redshifts are referred to the
CIV doublet.}
{\label{t4b}}
\epsfysize=26cm
\epsffile{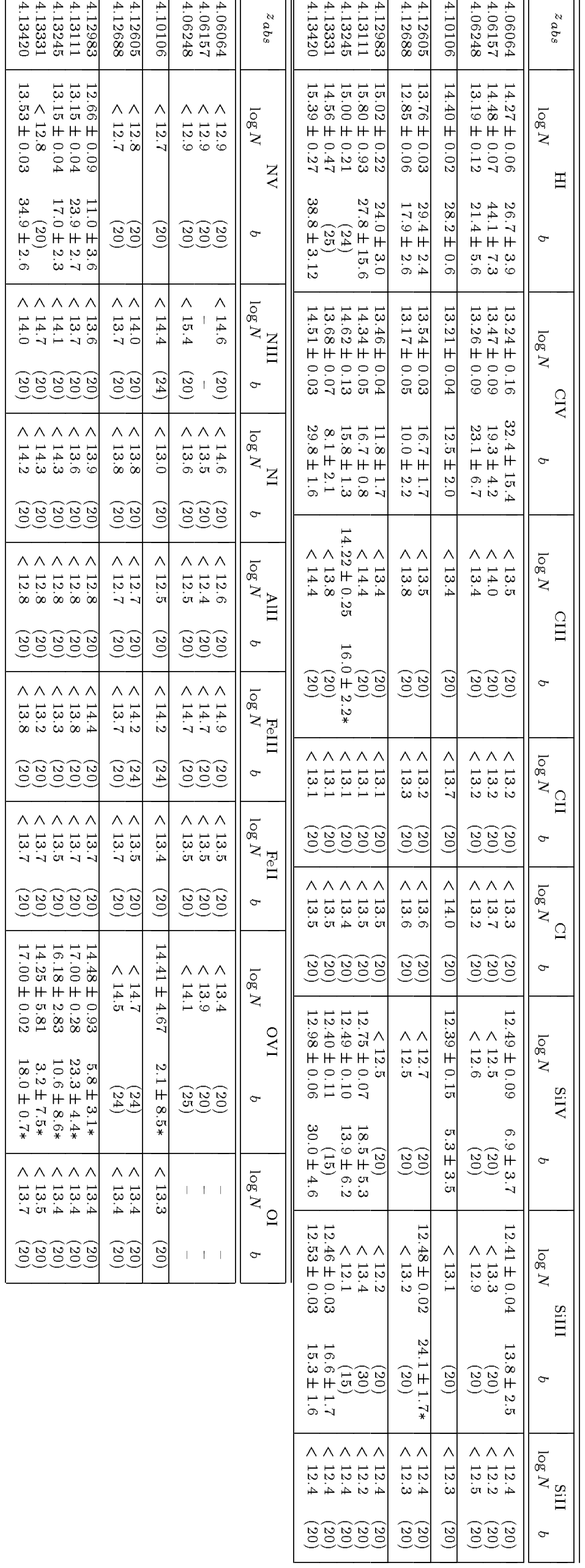}
\end{table*}

QSO absorption systems showing metal lines are interpreted as originating in 
intervening galaxies and thus represent an
important tool for the study of the chemical evolution of their
gaseous content.
The conversion of the observed column densities
to the metal content of an optically thin gas cloud is not
straightforward, since it depends on poorly known
parameters, mainly  the ionizing UV radiation 
 and  the cloud  geometry  and density.  
 An extensive discussion on the chemical evolution of galaxies
can be found in Timmes et al. (1995).  Abundance determinations are
traditionally reported in terms of an element abundance relative to
iron, [X/Fe], as a function of the iron--to--hydrogen ratio [Fe/H].  The
[Fe/H] ratio represents a chronometer in that the accumulation of iron
in the interstellar medium increases monotonically with time.  
Unfortunately in the high redshift metal absorption systems iron is
generally not observable, 
and abundances are derived respect to carbon 
and silicon.

Timmes et al. show that the [C/Fe] ratio is about constant
within a large range of metallicity. Observations of halo Galactic
stars  give [C/Fe] $\sim 0$ down to [Fe/H] $=-2$ (Wheeler et al.
1989).
For this reason
carbon can be used, neglecting any depletion by dust, as a tracer of
the chemical 
evolution of the absorbing clouds.  
Silicon is expected to be overabundant with respect to carbon by a factor
not higher than [Si/C] $=0.5-1~dex$ as a result of nucleosynthesis of 
massive metal poor stars. Relative silicon determinations in the 
interstellar medium have been 
presented by
Lu et al. (1996) for a sample of Damped \lya~systems. They found that
in the range of metallicity $-2.6<$ [Fe/H] $<-0.7$, there is a silicon
overabundance respect to iron in the range [Si/Fe] $\sim 0.2 - 0.6$.

\begin{figure}
\caption[1]{\label{fm} SiIV, CIV and HI absorption
for three intervening systems, one of which with two components.}
\epsfxsize=9cm
\epsfysize=9cm
\epsffile{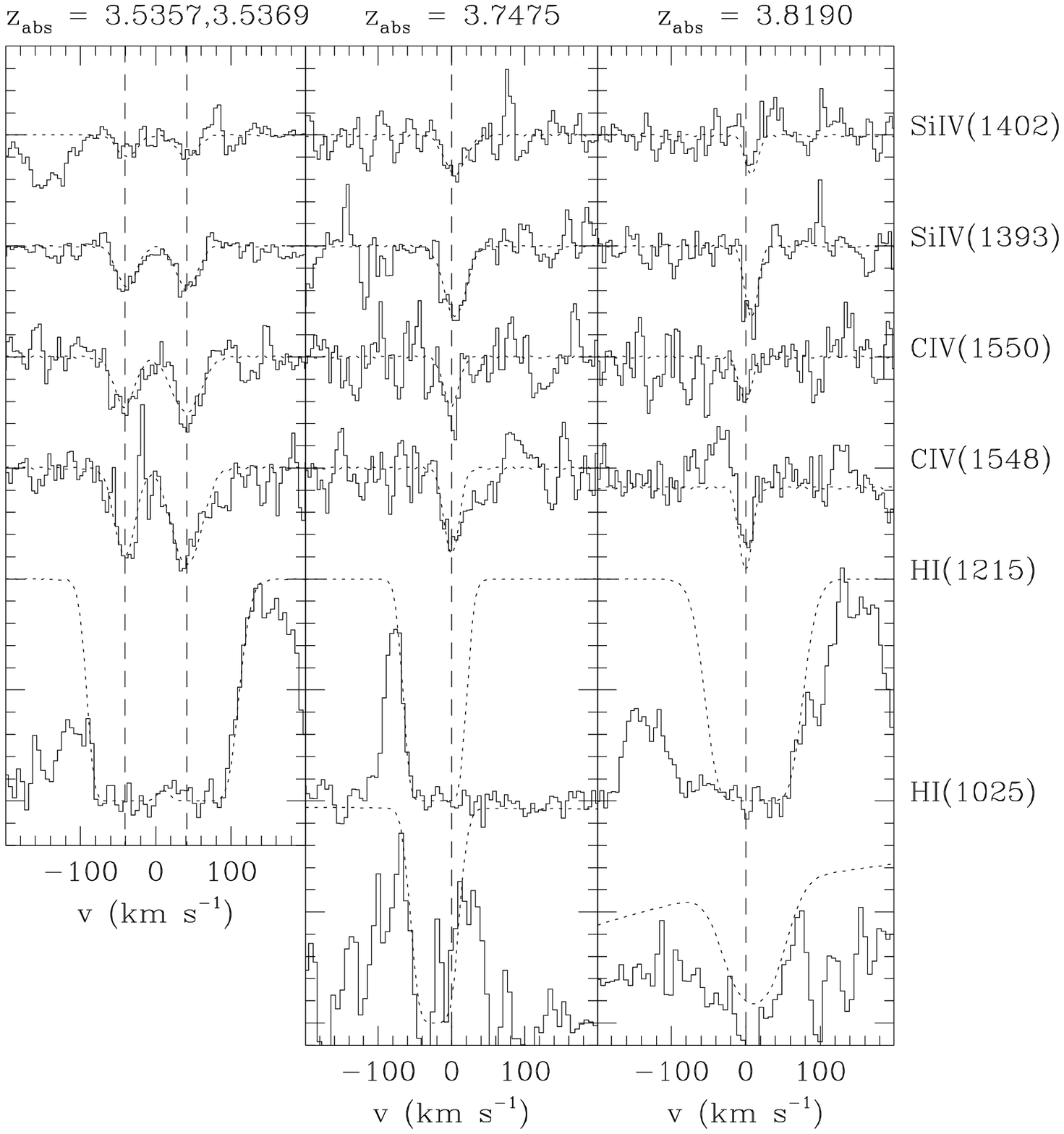}
\end{figure}

In this work we focus our attention to the metal systems which are
optically thin in hydrogen, 
similar to those studied by Cowie
et al. (1995), with HI column densities $\sim 10^{15}$ \cm2.

In three cases we detected SiIV together with CIV absorption
(Fig.~\ref{fm}).  Their
ratio provides an important information about the shape of the
ionizing UV background and the sources responsible for it
(Miralda-Escud\'e \& Ostriker 1990). In particular the ratio depends
on the average slope of the UVB around the HeII edge at 4 $Ryd$.  Both
model predictions and observations of far UV quasar spectra suggest
that the UVB shape beyond 1 $Ryd$ is more complex than a simple
power law because of HI and HeII absorption by the intergalactic
medium (Madau 1992).  An intrinsic steepening of the UVB at the HeII
edge could be also present if the ionizing sources are of stellar
origin.

In this discussion we considered all
absorption lines in every metal component 
having origin in a single--phase gas cloud, with a uniform density and
ionization state.

We assumed that the UV background radiation is the only ionizing
source. Thus the large $b$ values found in some system suggest the
presence of additional broadening mechanisms, like turbulent
broadening or the presence of several components, which cannot be
constrained given the limited $s/n$ and resolution.  Nevertheless, we
have verified that adding components or a turbulent broadening does not change 
significantly the total CIV and SiIV column densities, as expected for
unsaturated lines.

To estimate absolute and relative abundances for the three systems, we
used the standard photoionization code CLOUDY (Ferland 1991)
varying the most critical
parameters like the intensity and shape of the UVB and the total
density of the clouds.

\begin{figure}
\caption[1]{\label{fls} Portion of the long slit spectrum at low
resolution. Ticks show the position of Lyman limits for some of the
detected metal systems.}
\epsfxsize=8.5cm
\epsfysize=8.5cm
\epsffile{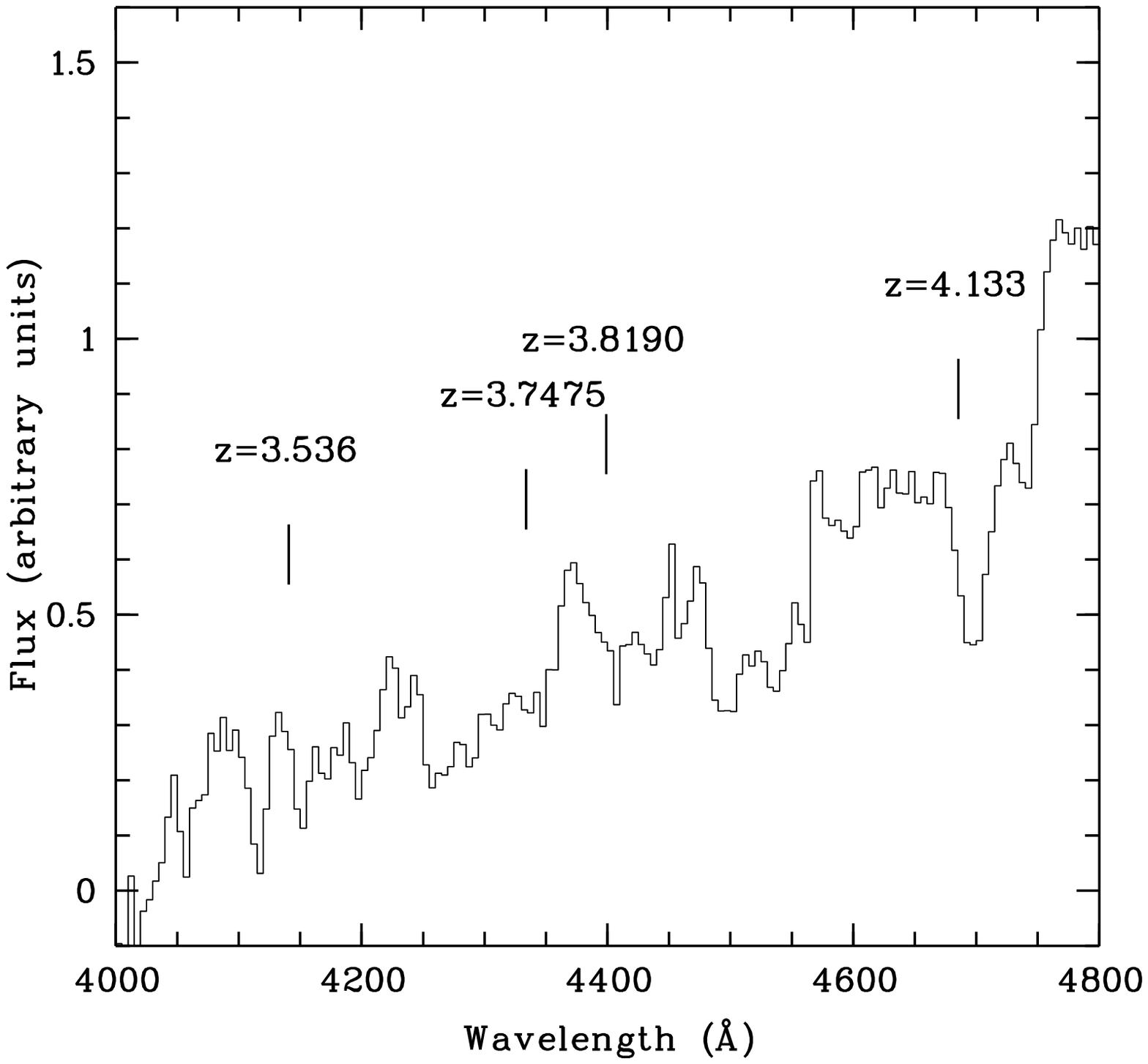}
\end{figure}

We considered two values for the UV flux at 912 \AA,
$J_{-22} = 1$ and
5.  For each of the four components at \za = 3.5357, 3.5369, 3.7475 and 3.8190
we have computed the metallicity [C/H] and the relative abundance
[C/Si] respect to solar values as a function of the total density
assuming different UVB shapes.  The HI column densities assumed
are those derived from the fit of the \lya~and/or \lyb~ shown in
Fig.~\ref{fm}. An upper limit can be estimated from the lack of the
Lyman limit edge (Fig.~\ref{fls}) to be $\log N_{HI} \lsim 16$.

\begin{figure*}
\caption[1]{\label{fm1} Metal content as function of the gas density
for three intervening systems (one of which with two components)
with HI, CIV and SiIV absorption. Here we assume $J_{-22} = 5$ and
varied $S_L$. Error bars for [C/H] due to the errors on the column
densities of HI and CIV are of about 
$0.2~dex$, $1~dex$, $0.5~dex$ and $0.3~dex$ for
the four systems at $z = 3.5357,~3.5369,~3.7475,~3.8190$ respectively.
Error bars for relative abundances of [C/Si] are typically of $0.1~dex$.}
\epsfxsize=18cm
\epsfysize=18cm
\epsffile{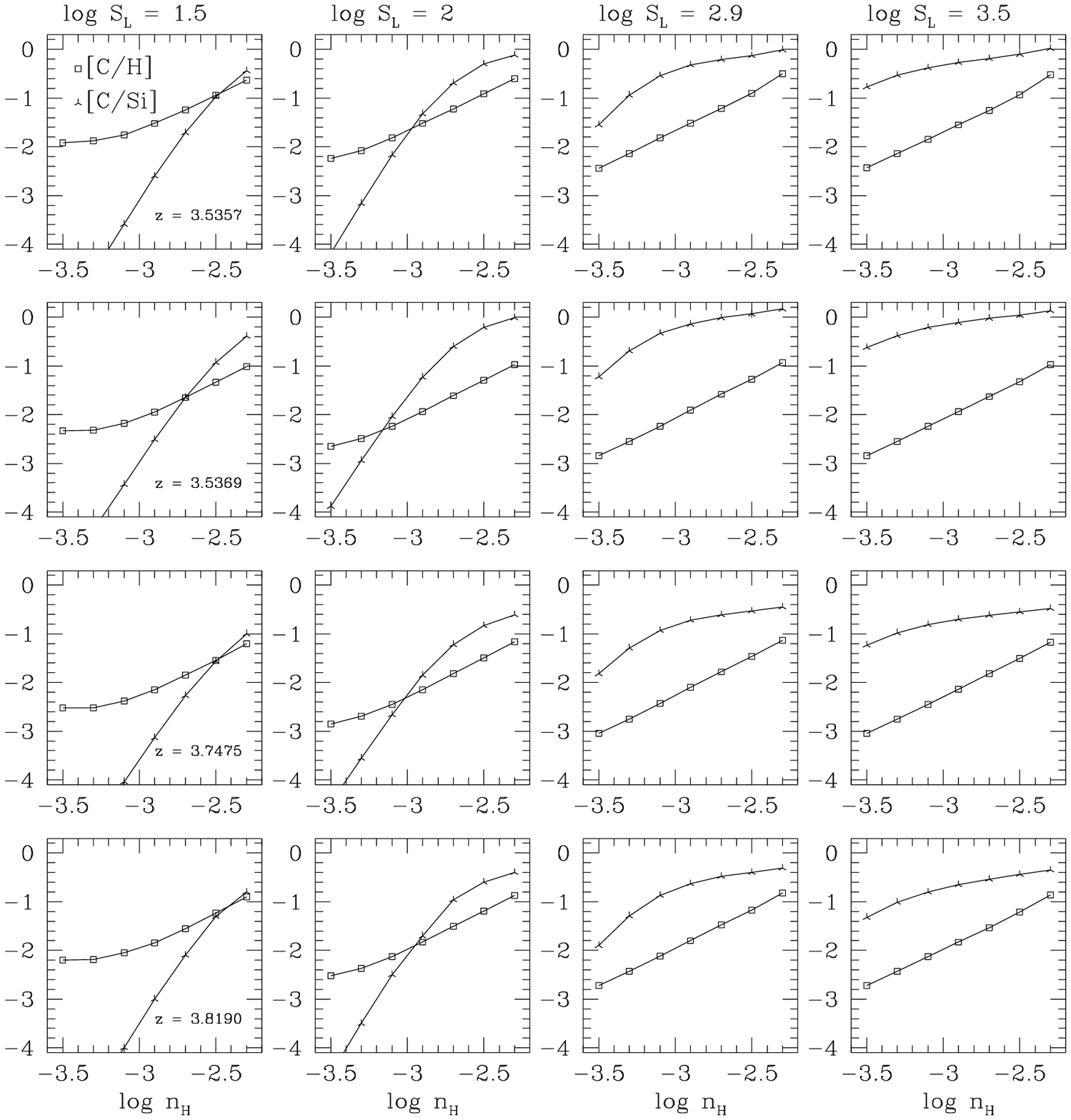}
\end{figure*}

Results are shown in Fig.~\ref{fm1}
for $J_{-22} = 5$ and different assumptions on the jump at the HeII
edge $S_L \equiv J(1Ryd)/J(4Ryd)$.  In the first plots from the left
the value $S_L = 25$ is assumed according to the predicted shape of
the UVB produced by quasars (Haardt \& Madau 1996).  In the other
cases, progressively higher values are assumed as expected 
when  stellar ionizing sources become the dominant
contributors.

In all the systems shown in Fig.~\ref{fm1} it appears that when  $\log S_L
<2$,  the [C/Si] ratio can be maintained within
acceptable values higher than $-1~dex$, in agreement with 
prediction by chemical
evolution models for galaxies (Timmes et al. 1995),
only for $\log n_H \gsim -2.5$. At such
high densities the metallicity is relatively high [C/H] $=
-1\sim-1.5$, while the cloud thickness is of only few hundreds of
parsec.
If we assume sizes at least one order of magnitude larger (few kpc) we
are forced to lower the density of the systems to $\log n_H = -3$ 
resulting in  an implausible overabundance of silicon over
carbon by 100--1000 times the solar value.

However, if we  assume a deeper UVB jump at the HeII edge we can obtain more
consistent results.  For $S_L \gsim 1000$ the silicon overabundance
is within a factor of 10 in all the systems considered and the
metallicities are about two orders of magnitude below the solar,
 while keeping the cloud size reasonably large.

Similar results hold also for $J_{-22} = 1$, where we have in general
higher values of [C/H] and [C/Si]. For $S_L \lsim 100$ we have [C/Si]
$\sim -1$ for $\log
n_H\sim -3$, but the sizes remain less than a kpc and the metallicities
would be unusually high, [C/H]$\gsim -1$, for these optically thin
clouds.

Errors on the abundance determinations are mostly systematic and
due to uncertainties on the model. However, uncertainties coming from
the fitting procedure are dominated by errors on the HI column
density, since these are much larger than for metal lines.  We have
verified that relative metal abundances do not change rescaling the HI
column density according to the upper and lower limit given by the
errors, while the metallicity do.

We notice that Songaila et al. (1995) have observed four metal systems
in the spectrum of 2000--33 deriving an average $S_L \sim 70$ at
$z=3.2$. Any further constraint on the shape of the UV flux coming
from the ionization state of low column density \lya~lines will be
important for the investigation of the evolution with redshift or
inhomogeneity of the ionizing UV population.

At this point we can only speculate that being the IGM highly ionized
at the Lyman limit as derived by the Gunn-Peterson test in the spectra
of the highest redshift quasars known (Giallongo et al. 1994), the UVB
should maintain a high intensity level beyond 1 $Ryd$ up to
$z=5$. Since the number density of quasars at redshift $z>3$ bends
down (Pei 1995), a change in the kind of ionizing population could
take place at these redshifts with a possible dominance of primeval
galaxies. Given the strong spectral difference at the HeII edge of the
two populations, an increasing jump at the HeII edge with increasing
redshift should be expected.

\subsection{The associated systems}

The abundances of the associated system have been determined in few
cases and in all of them high ionization with high metal content has
been derived (M{\o}ller et al. 1993; Wampler et al. 1993; Savaglio et
al. 1994; Petitjean et al. 1994). In all these measurements different
shapes have been assumed for the ionizing background, and this is
presumably dominated by the flux of the QSO itself.

Indeed the distance of the cloud cannot be derived in a
straightforward way because of the uncertainty in the measure of the
systemic redshift and because of the importance of the cloud
peculiar motions.

We found four associated systems, with a total of 11 components (Table
\ref{t4b}).  We estimate the metal abundances varying the intensity of
the ionizing source using different values of the ionization parameter
$U$. We assumed a simple power-law spectrum with spectral index
$\alpha = 1.5$ ($f_\nu \propto \nu^{-\alpha}$).  Also a value $\alpha
= 0.72$ has been considered. The first value is a reasonable
assumption for frequencies $\nu \geq 4~Ryd$ ($\lambda \leq 228$ \AA),
relevant for the considered ions, to keep the flux of the object at
low values at X--ray frequencies.  The second value is that used for
the proximity effect and is more appropriate for $\nu \leq 1~Ryd$
($\lambda \geq 912$ \AA).

\begin{figure*}
\caption[1]{\label{fm3} Metal content as function of the ionization
parameter for five associated systems. We assume $\log n_H = -2$ and
$\alpha=1.5$.
Errors bars for [C/H] due to the errors on the column
densities of HI and CIV
are of about $0.2~dex$, $0.06~dex$, $0.3~dex$, $1~dex$
 and $0.3~dex$ for
the five systems from the left to the right of the figure. Error bars for
relative abundances of [C/Si] and [N/C] are typically of $0.1~dex$.}
\epsfxsize=18cm
\epsfysize=10cm
\epsffile{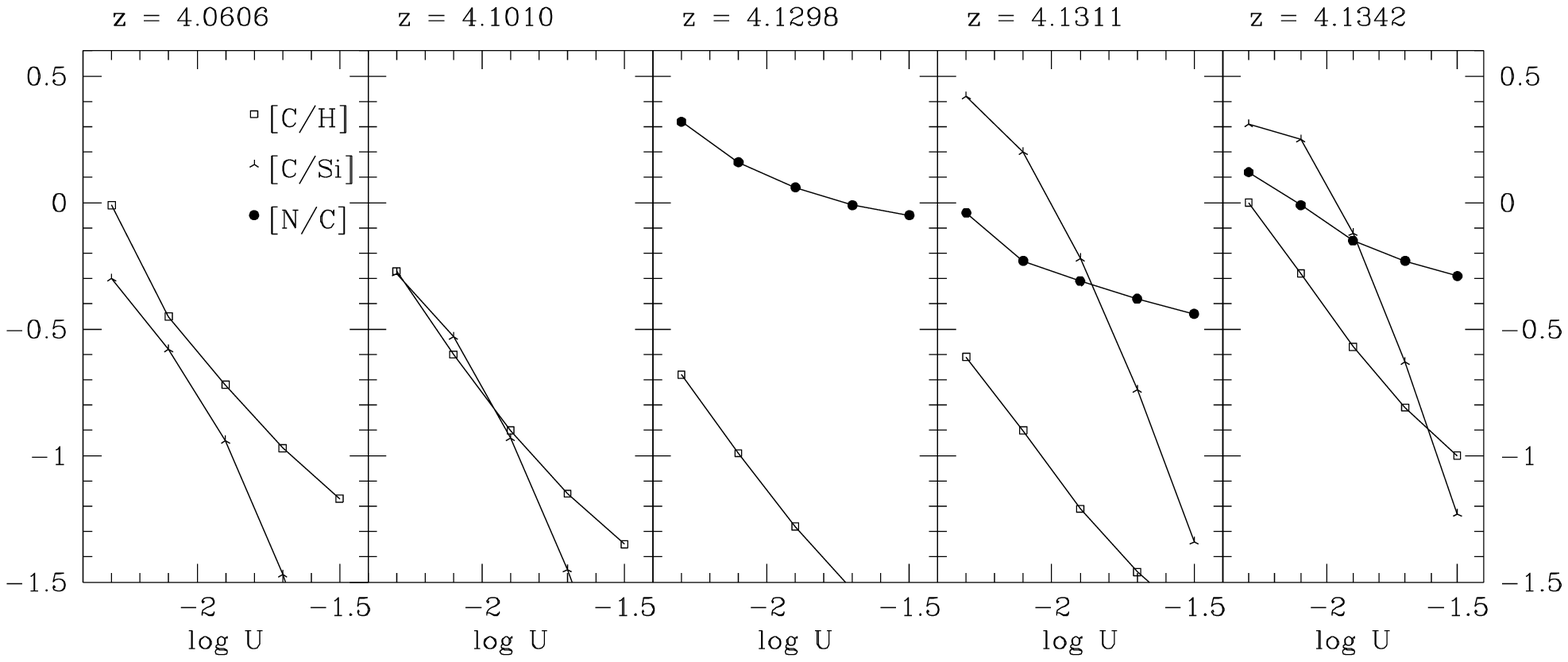}
\end{figure*}

As for the intervening systems, all the absorption lines in every
metal component are assumed to be originated in a single--phase gas
cloud, with a uniform density and ionization state.  The model results
are shown in Fig.~\ref{fm3} for five of the 11 measured components
where the presence of silicon and/or nitrogen provides some
constraints to the models.

For the strongest system at $z\sim4.13$ the metal content of the third
and fourth component is omitted because the relevant HI column
densities are particularly uncertain due to the saturated and blended
profile of the \lya~lines.  The \lyb~lines fall in the blue wing of
the damped \lya~line at \za = 3.39, making the line fitting more
complicated.  The total HI column density is $\log N_{HI} \simeq
16.2$.  From our grism spectrum we estimate an upper limit to the HI
optical depth at 912 \AA~ (Fig.~\ref{fls}), corresponding to a total
HI column density $\log N_{HI} \lsim 16$, consistent with the value
found from our best fit.  The OVI column densities reported in Table
\ref{t4b} are probably upper limits because of the confusion with the
\lya~forest and with the blue wing of the damped \lya~line at \za =
3.39. For the three components considered in this analysis,
results from the fit give ratios of the HI to OVI of $\log
(N_{HI}/N_{OVI}) \simeq -1.8, -3.9, -3.5$ respectively, suggesting the
OVI contamination.

Concerning the other three systems at lower redshifts, only the first
component of the system at $z\simeq 4.06$ shows SiIII and SiIV
together with CIV. The SiIV line is observed also in the system at
$z=4.1010$.  For the system at $z=4.126$ we only give for the first
component an upper limit to the SiIV abundance and a tentative SiIII
identification in the \lya~forest.

In Fig.~\ref{fm3} we show CLOUDY results for the selected components
of the associated systems. We report the metal abundances ([C/H]) and
the relative abundance of nitrogen and silicon respect to carbon as a
function of the ionization parameter $U$.  The adopted gas density is
$\log n_H = -2$ and the quasar spectral index is $\alpha=1.5$.

Constraints to the model can be derived from the upper limits on CIII
and from the detection of silicon.  The upper limit to the column
density of CIII gives a lower limit to the ionization parameter and an
upper limit to the metallicity.  The only exception is the $z=4.0606$
system for which we can only say that [C/Si] $>-1$ gives [C/H]
$>-0.7$ and that the observed CIV can be reproduced only for $\log U
> -2.4$

For the system at $z=4.1010$ the limits are $\log U > -2.3$ and [C/H]
$<-0.3$. If we require a relative silicon abundance lower than 1
$dex$, the metallicity is [C/H] $\simeq -1$ and $\log U < -1.9$.

In the system at $z=4.1298$, the non detection of CIII implies $\log U
> -1.9$ and consequently [C/H] $< -1.3$. The nitrogen relative
abundance remains close to that solar.

In the system at $z=4.1311$, the upper limit to the CIII column
density gives $\log U > -2.1$ and [C/H] $< -0.9$. The relative silicon
abundance is lower than 1 $dex$ if [C/H] $>-1.5$. For $-1.5<$ [C/H]
$<-0.9$, the relative nitrogen abundance is $-0.4 <$ [N/C] $<-0.2$. In
the last component, at $z=4.1342$, $\log U > -2.1$ and consequently
[C/H] $<-0.3$. From the relative silicon abundance we derive [C/H]
$>-0.9$. In this range of metallicity, the nitrogen abundance is
$-0.3<$ [N/H] $< 0.0$.

As for the intervening systems, errors on the HI column densities
are much larger than for metal lines and relative metal abundances do
not change rescaling the HI column density according to the upper and
lower limit given by the errors, while the metallicity do.

From the analysis with $\alpha=1.5$, we conclude that in the detected
metal systems the metallicity is undersolar, although values lower
than [C/H] $\sim-1$
are unlikely. This value is about one order of magnitude higher than that
found in intervening systems at about the same redshift.
A silicon overabundance seems favoured, while the
nitrogen abundance tends to be undersolar.

For $\alpha=0.72$, the values of [C/H] are higher with respect to
the previous analysis, up to 0.2 $dex$ for
the considered lower limit in the $\log U$ and of 
the order of 0.6 $dex$ for the upper limit of $\log U$. The [N/C]
values are slightly lower, while the 
Silicon overabundance results much higher, being almost in all cases 
higher by 1 $dex$.

Savaglio et al. (1994) reported slightly higher values for the
metallicity of the same associated systems.  The greater wavelength
coverage, the better $s/n$ of the present data and the new line
fitting procedure adopted, allow a more accurate estimates of the line
parameters.  Moreover it was possible to detect new absorption lines
and to infer more stringent upper limits to non--detected lines.  In
particular new upper limits on CIII absorptions have been used to
derive upper limits on the metallicity. In this work we also used a more
accurate photoionized code and we excluded the possibility of having
silicon overabundances higher by 1 $dex$,
lowering the upper limit to the metallicity.
Finally, we have focused our attention on those systems for which we
have reliable information only.

Metallicities derived for associated systems in $z_{em}=2-3$ quasars
show nearly solar values and in some cases  even higher than
solar (Wampler et al. 1993; Petitjean et al. 1994; M{\o}ller et
al. 1994). The lower metallicities derived for some systems at $z\simeq 4$
might be indicative of an evolution with redshift of the chemical abundances
in the associated systems, which however needs confirmation
with a larger sample of high quality data.

\section{Summary}

We have presented a list of absorption lines
observed in the spectrum of the 
quasar Q$0000-2619$ ($z_{em}=4.126$) with a
resolution of 13 \kms~and a signal--to--noise ratio of
$15-60$ per resolution element. The main results of the statistical
analysis can be summarized as follows:

\begin{itemize}
\item
The mode of the Doppler distribution for the \lya~lines is $\simeq 25$
\kms~with a dispersion of 7 \kms. The fraction of line with $10 < b
< 20$ \kms~is 17\%.
The Doppler values derived from uncontaminated \lyb~lines are smaller than
those obtained from the corresponding Ly$\alpha$, suggesting the
contribution of non saturated, non resolved components in the
\lya~profiles.

\item
On the basis of the proximity effect in this spectrum the integrated
UV background is estimated to be $J \sim 7 \times 10^{-22}$ erg
s$^{-1}$ cm$^{-2}$ ~Hz$^{-1}$ sr$^{-1}$, although only values of 
$J_{-22} < 4 $ and  $J_{-22} > 18$ are excluded at
more than $2\sigma$ level. This value is consistent with
previous estimates obtained at a lower $z$,
implying no appreciable redshift evolution of the 
UVB up to $z=4$,
in agreement with the absence of any Gunn-Peterson effect
up to $z=5$.

\item
The analysis of the intervening metal line systems has revealed in
particular the presence of three optically thin systems with $\log
N_{HI}\sim 15$ showing associated CIV and SiIV absorptions.  [Si/C]
ratios lower than 10 times the solar value
can be obtained only assuming a large jump in the spectrum of the
ionizing UV background beyond the HeII edge ($J_{912}/J_{228}\gsim
1000$). This result, if confirmed in other spectra at the same
redshift, is suggestive of a possible
increase of the stellar ionizing emissivity
over the declining quasar one for $z> 3$.

\item
The analysis of the associated metal line systems suggests abundances
generally below solar with typical values in the range
$0<$ [C/H]$< -1$. 
The derived values are lower than those estimated
for associated systems found in lower $z$ quasars.

\end{itemize}

\paragraph{Acknowledgements.}

It is a pleasure to thank M.~Limongi and G.~Marconi for useful remarks on an early
version of the paper. S.S. acknowledges the kind  hospitality at the 
Osservatorio Astronomico di Roma where most of this work was done.


\begin{thebibliography}{}

\bibitem[1986]{bes}
Bahcall J.~N., et al., 1996, ApJ, 457, 19. 
\bibitem[1986]{bes}
Bajtlik, S., Duncan, R. C., Ostriker, J. P. 1988, ApJ, 327, 570.
\bibitem[1986]{bes}
Bergeron J., Boiss\'e P., 1991, A\&A, 243, 344.
\bibitem[1986]{bes}
Bergeron J., Cristiani S., Shaver, P.~A., 1992,  A\&A, 257, 417.
\bibitem[1986]{bes}
Carswell R.~F., 1995,  {\it Proceedings of the ESO Workshop on Quasars Absorption Lines}, ed.~G.~Meylan, p313.
\bibitem[1986]{bes}
Charlton J.~C., 1995, {\it Proceedings of the ESO Workshop on Quasars Absorption Lines}, ed.~G.~Meylan, p405.
\bibitem[1986]{bes}
Chernomordik V.~V., 1995, ApJ, 440, 431.
\bibitem[1986]{bes}
Cowie L.~L., Songaila A., Kim T.-S., Hu E. M., 1995, AJ, 109, 1522.
\bibitem[1986]{bes}
Cristiani S., D'Odorico S., D'Odorico V.,Fontana A., Giallongo E., Savaglio S., 1996,  MNRAS, {\it submitted}.
\bibitem[1986]{bes}
Cristiani, D'Odorico, S., Fontana, A, Giallongo, E., Savaglio, S., 1995, MNRAS, 273, 1016.
\bibitem[1986]{bes}
Davidsen A.~F., Kriss G.~A., Zheng W., 1996, Nat, 380, 47.
\bibitem[1986]{bes}
D'Odorico, S. 1990, {\it ESO The Messenger}, 61, 51.
\bibitem[1986]{bes}
Espey, B. R. 1993, ApJL, 411, 59. 
\bibitem[1986]{bes}
Ferland G.~J., 1991, {\it OSU Astronomy Dept.~Internal Rept.}, 91-01
\bibitem[1986]{bes}
Fern\'andez--Soto A., Lanzetta K.~M., Barcons X., Carswell R.~F., Webb J.~K., Yahil A., 1996, ApJ, 460, L85.
\bibitem[1986]{bes}
Ferrara A., Giallongo E., 1996, MNRAS, {\it in press}.
\bibitem[1986]{bes}
Fontana A., Ballester P., 1995, {\it ESO The Messenger}, 80, 37.
\bibitem[1986]{bes}
Giallongo E., Cristiani, S., D'Odorico S., Fontana A., Savaglio S., 1996, ApJ, {\it in press}.
\bibitem[1986]{bes}
Giallongo, E., Cristiani, S., Fontana, A., Trevese, D. 1993, ApJ, 416, 137.
\bibitem[1986]{bes}
Giallongo, E., D'Odorico, S., Fontana, A., McMahon, R. G., Savaglio, S., Cristiani, S., Molaro, P., Trevese, D. 1994, ApJL, 425, L1.
\bibitem[1986]{bes}
Giallongo E., Petitjean P., 1994, ApJ, 426, L61.
\bibitem[1986]{bes}
Haardt F., Madau P., 1996, ApJ, 461, 20.
\bibitem[1986]{bes}
Hernquist L., Katz N., Weinberg D.~.H., Miralda--Escud\'e J., 1996, ApJ, 457, L51.
\bibitem[1986]{bes}
Hu E.~M., Kim T.-S., Cowie L.~L., Songaila A., Rauch M., 1995, AJ, 110, 1526.
\bibitem[1986]{bes}
Jakobsen P., Boksenberg A., Deharveng J.~M., Greenfield P., Jedrzejewski R.,  Paresce F., 1994, Nat, 370, 35.
\bibitem[1986]{bes}
Lanzetta, K.~M., Bowen, D.~V., Tytler, D., Webb, J.~K., 1995, ApJ, 442, 538.
\bibitem[1986]{bes}
Lu L., Sargent W.~L.~W., Barlow T.~A., 1996, Contribution to "Cosmic Abundances", the proceedings of the 6th Annual October Astrophysical Conference in  Maryland, {\it in press}.
\bibitem[1986]{bes}
Madau P., 1992, ApJ, 389, L1.
\bibitem[1986]{bes}
Miralda--Escud\'e J., Ostriker J.~P., 1990, ApJ,  350, 1. 
\bibitem[1986]{bes}
Miralda--Escud\'e J., Cen R., Ostriker J.~P., Rauch M., 1996, ApJ, {\it submitted}.
\bibitem[1986]{bes}
Molaro, P., D'Odorico, S., Fontana, A., Savaglio, S., Vladilo, G., 1995, A\&A, 308, 1. 
\bibitem[1986]{bes}
M{\o}ller P., Jakobsen P., Perryman M.~A.~C., 1993, A\&A, 287, 719.
\bibitem[1986]{bes}
Pei Y.~C., 1995, ApJ, 438, 623.
\bibitem[1986]{bes}
Petitjean P., Bergeron J., 1994, A\&A, 283, 759.
\bibitem[1986]{bes}
Petitjean P., Rauch M., Carswell R.~F., 1994, A\&A, 291, 29.
\bibitem[1986]{bes}
Rauch M., Carswell R.~F., Webb J.~K., Weymann R.~J., 1993, MNRAS, 260, 589.
\bibitem[1986]{bes}
Savaglio S., D'Odorico S., M{\o}ller P., 1994, A\&A, 281, 331.
\bibitem[1986]{bes}
Schneider, D.~P., Schmidt, M., Gunn, J.~E., 1989, AJ, 98, 1507.
\bibitem[1986]{bes}
Songaila A., Hu E.~M., Cowie L.~L., 1995, Nat, 375, 124.
\bibitem[1986]{bes}
Steidel C.~C., Dickinson M., Persson S.~E., 1994, ApJ, 437, L75.
\bibitem[1986]{bes}
Stone, R. P. S. 1977, Ap. J., 218, 767.
\bibitem[1986]{bes}
Stone, R. P. S., Baldwin, J. A. 1983, MNRAS, 204, 347.
\bibitem[1986]{bes}
Stone, R. P. S., Baldwin, J. A. 1984, MNRAS, 206, 241.
\bibitem[1986]{bes}
Timmes F.~X., Woosley S.~E., Weaver T.~A., 1995, ApJS, 98, 617.
\bibitem[1986]{bes}
Tytler D., Fan X.-M., Burles S., Cottrell L., Davis C., Kirkman D.,
Zuo L., 1995, {\it Proceedings of the ESO Workshop on Quasars Absoon Lines}, ed.~G.~Meylan, p289.
\bibitem[1986]{bes}
Wheeler J.~C., Sneden C., Truran J.~W., 1989, ARA\&A, Vol.~27, ed.~G.~Burbidge (Palo Alto: Annual Reviews), 279.
\bibitem[1986]{bes}
Walsh J.~R., 1992, {\it HST and New Oke Spectrophotometric Standard Stars--Flux Tables and Finding Charts}.
\bibitem[1986]{bes}
Wampler E.~J., Bergeron J., Petitjean P., 1993, A\&A, 273, 15.
\bibitem[1986]{bes}
Webb J.K., Parnell H.C., Carswell R.F., McMahon R.G., Irwin M.J., Hazard C., Ferlet R., Vidal--Madjar A. 1988, The ESO Messenger, 51, 15.
\bibitem[1986]{bes}
Williger G.~M., Baldwin J.~A., Carswell R.~F., Cooke A.~J., Hazard C., Irwin M.~J., McMahon R.~G., Storrie--Lombardi L., 1994, ApJ, 428, 574.

\end{thebibliography}
\end{document}